\documentclass[lettersize,journal]{IEEEtran}
\usepackage{amsmath,amsfonts}
\usepackage{algorithmic}
\usepackage{algorithm}
\usepackage{array}
\usepackage[caption=false,font=normalsize,labelfont=sf,textfont=sf]{subfig}
\usepackage{textcomp}
\usepackage{stfloats}
\usepackage{url}
\usepackage{verbatim}
\usepackage{cite}
\usepackage{xcolor}
\usepackage{graphicx}
\graphicspath{{./}{./figures/}}
\usepackage{placeins} 
\usepackage{hyperref}
\usepackage{authblk}

\usepackage{array}
\newcolumntype{L}[1]{>{\raggedright\let\newline\\\arraybackslash\hspace{0pt}}m{#1}}
\newcolumntype{C}[1]{>{\centering\let\newline\\\arraybackslash\hspace{0pt}}m{#1}}
\newcolumntype{R}[1]{>{\raggedleft\let\newline\\\arraybackslash\hspace{0pt}}m{#1}}

\begin{document}

\title{Randomness as Reference: Benchmark Metric for Optimization in Engineering}

\author[1]{Stefan Ivić}
\author[1]{Siniša Družeta}
\author[2]{Luka Grbčić}

\affil[1]{Faculty of Engineering, University of Rijeka}
\affil[2]{Applied Math and Computational Research, Lawrence Berkeley National Laboratory}


\maketitle

\begin{abstract}
Benchmarking optimization algorithms is fundamental for the advancement of computational intelligence. However, widely adopted artificial test suites exhibit limited correspondence with the diversity and complexity of real-world engineering optimization tasks. This paper presents a new benchmark suite comprising 235 bounded, continuous, unconstrained optimization problems, the majority derived from engineering design and simulation scenarios, including computational fluid dynamics and finite element analysis models. In conjunction with this suite, a novel performance metric is introduced, which employs random sampling as a statistical reference, providing nonlinear normalization of objective values and enabling unbiased comparison of algorithmic efficiency across heterogeneous problems. Using this framework, 20 deterministic and stochastic optimization methods were systematically evaluated through hundreds of independent runs per problem, ensuring statistical robustness. The results indicate that only a few of the tested optimization methods consistently achieve excellent performance, while several commonly used metaheuristics exhibit severe efficiency loss on engineering-type problems, emphasizing the limitations of conventional benchmarks. Furthermore, the conducted tests are used for analyzing various features of the optimization methods, providing practical guidelines for their application. The proposed test suite and metric together offer a transparent, reproducible, and practically relevant platform for evaluating and comparing optimization methods, thereby narrowing the gap between the available benchmark tests and realistic engineering applications.
\end{abstract}


\FloatBarrier

\section{Introduction}

Rigorous evaluation of optimization algorithms is fundamental to advancing computational intelligence. Progress in black-box optimization—where the objective function is unknown or inaccessible—depends on representative benchmarks. Yet many existing suites fail to capture the complexity of real engineering problems, creating a disconnect between test-suite performance and practical effectiveness.

The No Free Lunch theorem \cite{wolpert2002no} asserts that no single algorithm is universally superior, making it essential to identify which algorithms excel for specific problem types or function landscapes. This calls for a new generation of benchmarks that reflect real-world engineering design, offering a more accurate and challenging platform for evaluating future algorithms.

A second cornerstone of rigorous benchmarking is the metrics used to evaluate performance. While benchmark suites define the problems, metrics establish how outcomes are interpreted. Conventional measures such as convergence curves or best-found objective value are often insufficient for expensive engineering optimization, where robustness, efficiency, and problem conditioning can be as critical as accuracy.

To bridge this gap, we propose a benchmark suite that mirrors real-world engineering optimization scenarios, providing a reliable tool for evaluating leading optimization methods and their many recent variants. The framework targets real-variable, bounded, unconstrained engineering problems. We also introduce novel evaluation methodologies that move beyond traditional metrics, offering deeper insights into algorithmic behavior and practical utility.

We implement this framework using Indago, an open-source Python module for numerical optimization developed and maintained by the authors since 2019. Indago provides a collection of modern methods for optimizing real fitness functions over continuous domains and, though initially built for in-house research and teaching, has since found wide application across engineering fields.

Many methods used here are provided by Indago but are not exclusive to it; they are among the most common stochastic optimization techniques, making them well suited to this study. To broaden the evaluation, we also include algorithms from other well-established implementations.

\section{Literature overview}

The origins of algorithm benchmarking trace back to classical, mathematically defined test functions such as Rastrigin, Rosenbrock, and Sphere \cite{jamil2013literature, naser2025review}. Known as artificial landscapes, these functions provide controlled environments for systematically assessing fundamental algorithmic properties, including convergence rate, accuracy, robustness, and efficiency. The rationale is that exposing an algorithm to a broad set of such landscapes reveals its relative strengths and limitations across diverse topological challenges.

Constructing mathematical test functions is inherently an act of abstraction that strips away many of the complexities and irregularities of real-world problems, yielding simplified—and sometimes misleading—representations. This introduces several intrinsic limitations. Many widely used functions exhibit special mathematical structures unrepresentative of general optimization problems. The Rosenbrock function, for example, is formulated as a nonlinear least-squares problem with a zero-valued global optimum, a structure exploitable by specialized methods such as the Gauss–Newton algorithm—advantages that may not carry over to more general problems.

A more fundamental limitation is the common practice of placing the global optimum at or near the center of the search space (e.g., the zero vector) \cite{kudela2022critical}. Though seemingly benign, this choice has distorted the field, enabling numerous purportedly "novel" metaheuristics that are not genuinely innovative but instead exploit a center bias through explicit or implicit "check-the-middle" strategies. Predisposed to converge toward the center of the feasible domain, such algorithms achieve artificially strong results on these flawed benchmarks. When the optimum is shifted away from the origin, their performance often deteriorates sharply, exposing a lack of general problem-solving capability \cite{kudela2023evolutionary, kudela2022commentary, castelli2022salp, deng2024deficiencies, rajwar2025structural}.

As evolutionary computation matured, the community moved from ad-hoc use of classic test functions toward standardized frameworks, most notably the Black-Box Optimization Benchmarking (BBOB) suite within the Comparing Continuous Optimizers (COCO) platform \cite{hansen2010comparing, hansen2010real, elhara2019coco, hansen2021coco} and the competition-driven Congress on Evolutionary Computation (CEC) benchmarks \cite{CEC2011, CEC2013, CEC2017}. Both introduce rotations and translations to avoid center bias. COCO provides a rigorous, automated environment for evaluating black-box optimizers, with structured function groups probing separability, conditioning, and multimodality, plus extensions for noisy, constrained, mixed-integer, large-scale, and bi-objective problems \cite{hansen2010comparing, varelas2018comparative, brockhoff2022using}. The CEC competitions, in parallel, established evolving suites of unimodal, multimodal, hybrid, and composition functions, serving as community-wide benchmarks and fostering rapid innovation through competitive testing.

Despite their rigor and influence, both BBOB and CEC lack the diversity and constraint complexity typical of real-world engineering problems \cite{vskvorc2020understanding, munoz2020generating, christie2018investigating}. Exploratory Landscape Analysis studies show these suites cover only a narrow portion of possible problem types, leading to poor generalization and algorithm overfitting \cite{munoz2020generating, lacroix2019limitations, vskvorc2020understanding, lang2021exploratory, kudela2022new, ibehej2025investigation}. Algorithmic superiority demonstrated on them may thus reflect artifacts of benchmark design rather than genuine robustness, calling for a shift toward more representative, application-driven benchmarks grounded in real-world complexity.

In response, a growing body of work has focused on domain-specific benchmark suites that embed the structure, constraints, and objectives of real engineering and computational systems, offering a more faithful test of practical capability. Examples include CATBench for compiler autotuning, which captures discrete, conditional, and permutation parameter spaces with realistic performance metrics \cite{torring2024catbench}, and VRPBench, which replaces artificial Euclidean instances with large-scale, city-derived routing graphs and multi-objective logistics goals \cite{zeni2016vrpbench}. Engineering domains contribute rich, high-dimensional suites: aerodynamic and CFD-driven shape optimization exposes hundreds of continuous design variables and costly evaluations \cite{lyu2015aerodynamic, nadarajah2015adjoint}, structural truss optimization enforces nonlinear stress/displacement constraints \cite{sankaranarayanan1994truss}, and canonical tasks such as pressure-vessel sizing tie objectives to manufacturing and code-based safety constraints \cite{sandgren1990nonlinear, gandomi2011benchmark}. Further examples span robotics motion planning \cite{kuudela2023collection, liu2022benchmarking}, uncrewed aerial vehicle path planning \cite{shehadeh2025benchmarking}, solar power plant simulation \cite{andres2025solar}, and simulation-optimization problems \cite{eckman2023simopt}. Collectively, these suites show how realistic variable types (mixed continuous/discrete/permutation), black-box or hidden constraints, and multi-objective criteria produce problem features that classical mathematical functions fail to capture.

Despite their value, existing domain-specific benchmarks remain fragmented. Each targets a specific application area with its own data formats, assumptions, and evaluation methods, hindering cross-domain comparison. Common issues include reliance on costly simulators or proprietary tools, inconsistent reporting of problem difficulty and evaluation limits, and limited inclusion of conditional or hierarchical parameters.

Building on these observations, we developed IndagoBench25, a benchmark suite designed to mirror real-world engineering optimization scenarios and evaluate the practical capabilities of modern algorithms. It comprises 235 functions, most derived from engineering applications and a substantial portion simulation-based, with Python-based implementations and interfaces spanning problems from analytical formulations to those using external simulation tools. Built on open-source code as a shared foundation, it ensures high availability and reproducibility. IndagoBench25 is paired with new evaluation methodologies that go beyond traditional performance metrics, using random sampling as a reference to provide deeper insights into algorithmic behavior and utility.

\section{Methodology}

The analysis and comparison of optimization approaches, whether through changes in the optimization problem or method, fundamentally rely on the value of the function being minimized. The fitness or cost values associated with solving engineering problems can vary significantly, making their convergence dynamics challenging to interpret. As a solution for this, normalizing fitness values have been proposed in the literature \cite{hansen2010real}. Building on this foundation, we propose a practical and objective method for normalizing values produced by the objective function.

For this purpose, we define a metric denoted as $\mathbb{G}$ for nonlinear normalization based on three objectively defined reference points. The first reference point is the solution of the optimization problem at hand, or more precisely, the best-known solution since finding an exact solution is often difficult for many engineering problems. We assign $\mathbb{G} = 1$ to this solution reference point, which is denoted as $f^-$.
The second reference point relies on the basic spatial statistics of the objective function. It is the median of objective values obtained from uniformly randomly sampling the search space, which we denote as $f^+$. The isoline of this median value effectively splits the search space into two equally sized parts, representing the expected fitness of a single, completely uninformed guess. This reference point is assigned $\mathbb{G} = -1$.
The third reference point evaluates the performance of a naive, unbiased search over a fixed evaluation budget. Although being inefficient, it represents a naive baseline optimization method which we refer to as Random Search (RS). Unlike the single-sample approach used for the second reference point, RS tracks the best-found solution over a predetermined allowance of evaluations equal to the computational budget given to all other tested optimization methods. The median of these best-found solutions across multiple independent RS runs is used as the third reference point, denoted as $f^\circ$, and is associated with $\mathbb{G} = 0$. In summary, while $f^+$ represents the expected outcome of a single random evaluation, $f^\circ$ represents the expected outcome of utilizing the entire available computational budget purely on random guessing.

We aim to define a nonlinear normalization metric $\mathbb{G}$ for the bounded unconstrained minimization problem
\begin{equation}
	\min_{\mathbf{x}\in\Omega} ~ f(\mathbf{x})~,
	\label{eq:minimization_problem}
\end{equation}
where $\mathbf{x}\in\mathbb{R}^n$ is the optimization vector, $f:\mathbb{R}^n\to\mathbb{R}$ is the objective function and $\Omega$ is $n$-dimensional bounded real-valued search space. Also, $\mathbf{x}_{lb}\in\mathbb{R}^n$ and $\mathbf{x}_{ub}\in\mathbb{R}^n$ are lower and upper bounds, respectively, used in the element-wise inequality constraints that define the search space:
\begin{equation}
	\mathbf{x}_{lb} \leq \mathbf{x} \leq \mathbf{x}_{ub}.
\end{equation}

First we perform a linear normalization between the solution reference point and the random sampling median reference point. The best found solution of \eqref{eq:minimization_problem} is $\mathbf{x}^-\in\Omega$ which ensures the minimal value of objective function
\begin{equation}
	f^-=f(\mathbf{x}^-) \leq f(\mathbf{x}), \quad \forall \mathbf{x}\in\Omega ~~\text{of all found solutions}.
\end{equation}
In contrast, random sampling obtains sub-optimal results with higher objective values, hence we denote their median value with $f^+$.
Normalized linear grade $\rho$ scales then the objective value $f$ from the range $[f^-, f^+]$ to unit range $[0, 1]$:
\begin{equation}
	\rho(f) = \frac{f - f^-}{f^+ - f^-}.
	\label{eq:rho}
\end{equation}
Note that $\rho$ mapping is not bounded to $[0, 1]$ since at least half of the domain has $\rho>1$. Finding a new best solution below $f^-$ will produce $\rho<0$ if $f^-$ is not updated. Since we update $f^-$ every time a new best solution was found while conducting the benchmark on a certain problem, this does require reevaluation of all $\rho$ values.

Finally, we introduce non-linearity to the mapping by adding the third reference point $f^\circ$. With sufficient statistical samples, one can reasonably expect that the three reference points are distinct and ordered, $f^- < f^\circ < f^+$. Since this ordering is ensured, we can require that a non-linear mapping is monotonic decreasing, i.e. having a property of $\mathbb{G}(f_a) > \mathbb{G}(f_b)$ if $f_a < f_b$.
All this can be satisfied by using a normalized logarithmic map:
\begin{equation}
	\mathbb{G}(f) = 1 - 2 \frac{\log\left(\rho(f) \cdot (\alpha - 1) + 1\right)}{\log \alpha}~,
	\label{eq:g_formulation}
\end{equation}
with yet unknown degree of freedom controlled by unknown parameter $\alpha$. This way the metric $\mathbb{G}$ employs a logarithmic scale mapped to $[-1, 1]$, with the capability to extend beyond this range in case a solution better than $f^-$ is found. As long as no better solution than $f^-$ is found, it robustly satisfies $\mathbb{G}(f^-) = 1$ by having the numerator equal to zero ($\rho(f^-)=0$), and $\mathbb{G}(f^+)=-1$ by having the numerator equal to the denominator ($\rho(f^+)=1$).

We have yet to decide on the value of $\alpha$, by considering the third reference point $\mathbb{G}(f^\circ)=0$. Using this point, the equation \eqref{eq:g_formulation} can be solved for $\alpha$:
\begin{equation*}
	1 - 2 \frac{\log\left(\rho^\circ \cdot (\alpha - 1) + 1\right)}{\log \alpha} = 0.
\end{equation*}
This simplifies to 
	$\rho^\circ(\alpha-1) + 1 = \sqrt{\alpha}$,
which, using substitution $\beta=\sqrt{\alpha}$, yield a quadratic problem:
	$\rho^\circ\beta^2 - \beta + 1 - \rho^\circ$.
The solutions for $\beta$ are
\begin{equation*}
	\beta=\frac{1\pm\sqrt{1 + 4\rho^\circ(\rho^\circ - 1)}}{2\rho^\circ}
\end{equation*}
and by examining the solution set, the suitable root is chosen, leading to the formula for $\alpha$:
\begin{equation}
	\alpha=
	\begin{cases}
		\left(\frac{1 + \sqrt{1 + 4\rho^\circ(\rho^\circ - 1)}}{2\rho^\circ}\right)^2  & \text{ for } \rho^\circ<0.5\\
		\left(\frac{1 - \sqrt{1 + 4\rho^\circ(\rho^\circ - 1)}}{2\rho^\circ}\right)^2  & \text{ for } \rho^\circ>0.5~.\\
	\end{cases}
	\label{eq:alpha_formulation}
\end{equation}

The unambiguous $\alpha$ value for any observed function, allows for the calculation of $\mathbb{G}$ for any objective value $f\in[f^-, f^+]$. 

Expressions \eqref{eq:g_formulation} and \eqref{eq:alpha_formulation} fail at a singularity point $\rho^\circ=0.5$ for which mapping should be linear and $\alpha=1$. We bridge this discontinuity by employing $f \to \mathbb{G}$ map calculated as
\begin{equation*}
	\mathbb{G}(f) = 1 - 2\cdot\rho(f) \quad\text{ if } |\rho^\circ - 0.5| < \epsilon~.
\end{equation*}
We use $\epsilon=10^{-3}$ for all calculations.

\begin{figure}[h!]
	\centering
	\includegraphics[width=1\linewidth, trim={1mm 1mm 0mm 0mm}, clip]{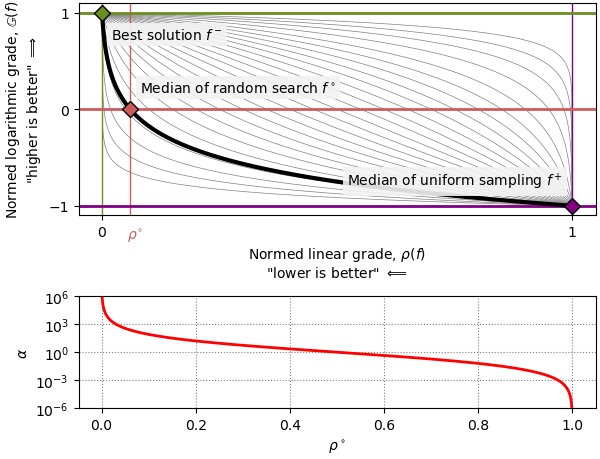}
	\caption{Top plot shows an example of $\mathbb{G}$ mapping and the possible relations of the normalized logarithmic grade $\mathbb{G}$ to the normalized linear grade $\rho(f)$. Bottom plot indicates the sensitivity of $\alpha$, and consequently $\mathbb{G}$, to limiting values of $\rho$, which happens when $f^\circ$ gets too close to $f^-$ or $f^+$.}
	\label{fig:lin_log_relation}
\end{figure}

A demonstration on how the reference points define the $\mathbb{G}(\rho)$ map is given in the top plot of Figure~\ref{fig:lin_log_relation}, with the bottom plot showing the relationship between $\alpha$ and $\rho^\circ$. Logarithmic mapping for the $\mathbb{G}$ metric is appealing because optimization progress typically slows near the optimum, and $\log$ scale makes these diminishing gains more visible. It also explicitly accommodates fitting through the three reference points while maintaining smooth and  monotonic behavior.

It should be noted that, in the rare case of discovering a new global optimum $f^-$ after $\mathbb{G}(f)$ has been already defined as per the previously attained $f^+$, $f^\circ$, and $f^-$ values, the $\mathbb{G}(f)$ transformation does not necessarily have to be updated immediately. The transformation itself would still be calculable and would yield $\mathbb{G} > 1$ values, which can still be compared to and averaged with other, "regular" $\mathbb{G} < 1$ values.

The choice of a logarithmic transformation for the $\mathbb{G}$-metric is mathematically motivated by the diminishing returns typical of optimization algorithms, effectively stretching critical late-stage refinements to evaluate accuracy where it matters most. Conceptually, this relates to the widely used COCO framework \cite{hansen2021coco, hansen2010comparing}, which measures the runtime to reach predefined, logarithmically spaced quality targets. COCO favors this fixed-target approach, arguing that raw objective values obtained under a fixed evaluation budget lack interpretability across diverse problems. However, while effective for analytical test suites, fixed-target assessments are often impractical for computationally expensive, simulation-based engineering problems where rather strict evaluation budgets must be enforced.

The proposed $\mathbb{G}$-metric provides a continuous, fixed-budget counterpart to this philosophy. Rather than relying on discrete, predefined targets that may be arbitrary across heterogeneous engineering domains, it dynamically calibrates a continuous logarithmic curve using the data-driven performance of RS ($f^\circ$). This normalization successfully maps fixed-budget outcomes into a dimensionless, interpretable scale inherently adjusted to the baseline difficulty of the problem.

\section{Optimization methods}

The proposed metric was applied on the IndagoBench25 suite, on which a range of 20 modern well-known optimization methods were tested. These include deterministic and stochastic methods, as well as local and global search methods. Considering that many real-world engineering problems are typically distinctly global, the selection of methods used is biased towards methods with strong global search capabilities.

Besides the 20 selected methods, a representative of surrogate-based Bayesian optimization methods, namely the Tree-structured Parzen Estimator (TPE) \cite{TPE}, was also intially considered. However, the computational overhead of TPE has shown to be substantial and effectively prohibitive for IndagoBench25 problems. For this reason TPE was excluded from the analysis and the incomplete TPE results are provided only as part of the supplementary material.

The use of local search methods within a global optimization context warrants clarification. Although local optimization methods are in principle unfit for solving global problems, in a real-world optimization scenario a fixed allowance of function evaluations can typically be used either for a smaller number of global optimization runs, or for many local optimization attempts initiated from random starting solutions. The latter strategy can sometimes be very successful, as shown in the benchmark results presented in this paper. 
	
An overview of the methods, their parameters and used software implementations is given in Table \ref{tab:methods}. More information is available in the supplementary material (see Section \ref{sec:supplementary}). 

\begin{table*}
	\scriptsize
	\centering
	\caption{Used optimization methods}
	\begin{tabular}{L{57mm}cccL{40mm}cc}
		\hline
		\textbf{Method} & \textbf{Acronym} & \textbf{Scope} & \textbf{Type} & \textbf{Parameters} & \textbf{Software} & \textbf{Ref.} \\
		\hline
		Random Search & RS & global & stochastic & $size_{batch} = D$ & Indago 0.5.4 & - \\
		\hline
		Limited-Memory Broyden-Fletcher-Goldfarb- Shanno Algorithm with Bounds & LBFGSB & local & deterministic & $eps=0.1(ub - lb)$ & Scipy 1.11.1  & \cite{BFGS} \\
		\hline
		Nelder-Mead & NM & local & deterministic & $step_{init} = 0.4$  & Indago 0.5.4  & \cite{GaoHanNM} \\
		\hline
		Multi-Scale Grid Descent & MSGD & local & deterministic & $divisions=10$; $base=4$; $scale_{max}=15$  & Indago 0.5.4  & - \\
		\hline
		Sequential Randomized Coordinate Shrinking & SRACOS & global & stochastic & $n_{sre}=5$; $\lambda=0.01$; $\alpha=0.5$ & ZOOpt 0.4.2  & \cite{ZOOpt} \\ 
		\hline
		Dual Annealing & DA & global & stochastic & $q_v=2.62$; $q_a=-5$; $T_0=5230$; $\rho=2\cdot10^{-5}$ & Scipy 1.11.1  &  \cite{DA1} \\
		\hline
		Controlled Random Search & CRS & global & stochastic & $n=10(D+1)$ & NLopt 2.7.1  & \cite{CRS} \\
		\hline
		Stochastic Global Optimization & STOGO & global & stochastic & - & NLopt 2.7.1  & \cite{STOGO} \\
		\hline
		Electromagnetic Field Optimization & EFO & global & stochastic & $n = \max (50, D)$; $R_{rate} = 0.25$; $Ps_{rate} = 0.25$; $P_{field} = 0.075$; $N_{field} = 0.45$ & Indago 0.5.4  & \cite{EFO} \\
		\hline
		Particle Swarm Optimization & PSO & global & stochastic & $n = \max (10, D)$; $w = 0.72$; $c_1 = 1$; $c_2 = 1$ & Indago 0.5.4 &  \cite{PSO} \\
		\hline
		Artificial Bee Colony & ABC & global & stochastic & $n = \max (10, 2D)$; $t_{limit} = n D / 2$ & Indago 0.5.4  & \cite{ABC} \\
		\hline
		Squirrel Search Algorithm & SSA & global & stochastic & $n = \max (20, 2D)$; $ata = 0.5$; $p_{pred} = 0.1$; $c_{glide} = 1.9$  & Indago 0.5.4 & \cite{SSA} \\
		\hline
		Bat Algorithm & BA & global & stochastic & $n = \max (15, D)$; $loud = 1$; $p_{rate} = 0.001$; $\alpha = 0.9$; $\gamma = 0.1$  & Indago 0.5.4  & \cite{BA} \\
		\hline
		Grey Wolf Optimizer & GWO & global & stochastic & $n = \max (10, D)$  & Indago 0.5.4  & \cite{GWO} \\
		\hline
		Manta Ray Foraging Optimization & MRFO & global & stochastic & $n = \max (10, D)$; $f_{som} = 2$ & Indago 0.5.4  & \cite{MRFO} \\
		\hline
		Fireworks Algorithm & FWA & global & stochastic & $n = D$; $m_1 = D/2$; $m_2 = D/2$  & Indago 0.5.4 &  \cite{FWA} \\
		\hline
		Successful History-Based Adaptive Differential Evolution with Linear Population Size Reduction  & LSHADE & global & stochastic & $n_{init} = \max (30, 5 D)$; $f_{archive} = 2.6$; $h_{size} = 6$; $p_{mut} = 0.11$  & Indago 0.5.4  & \cite{LSHADE} \\
		\hline
		Genetic Algorithm & GA &  global & stochastic & $n=2D$; $p_x=0.5$; $p_m=0.9$, $k_{tour}=2$ &  pymoo 0.6.1.3 & \cite{pymoo} \\
		\hline
		CH Evolutionary Strategy & ESCH & global & stochastic & $n=10(D + 1)$ & NLopt 2.7.1  &  \cite{ESCH} \\
		\hline
		Covariance Matrix Adaptation Evolution Strategy & CMAES & global & stochastic & $n=4+\text{int}(3 \log D)$; $\sigma=0.5$ & pymoo 0.6.1.3 &  \cite{CMAES} \\
		\hline
	\end{tabular}
	\label{tab:methods}
\end{table*}

\section{Benchmark test}

In tandem with the proposed $\mathbb{G}$-metric, IndagoBench25 was developed as an attempt to provide a "one stop shop" for testing optimization methods' capabilities, robustness and, most importantly, their suitability for solving various types of engineering optimization problems. 

IndagoBench25 is a test suite comprised of 235 optimization functions with dimensionality $D \in [3, 58]$, as summarized in the Table \ref{tab:IndagoBenchFuns}. 117 of these functions are based on engineering problems, out of which 34 are simulation based, meaning a numerical model of some kind is used for computing their function value. The numerical models utilized are models typically used for engineering purposes, namely various computational fluid dynamics (CFD) models and finite element method (FEM) based structural analysis models.

\begin{table}[h]
	\centering
	\scriptsize
	\caption{Benchmark test functions overview}
	\begin{tabular}{lccc >{\centering\arraybackslash}b{5mm}}
		\hline
		\textbf{Function family} & \textbf{Abbrev.} & \textbf{Size} & $D$ & \textbf{Sim. based} \\ \hline
		CEC 2014 test functions  & CEC  & 90 & $[10, 50]$ & $\times$ \\ 
		Analytical engineering problems  & AEP  & 19 & $[3, 36]$ & $\times$ \\ 
		Empirical regression  & ER  & 11 & $[3, 9]$ & $\times$  \\ 
		Shortest path problems  & SP  & 24 & $[5, 50]$ &$\times$ \\ 
		Ergodic coverage  & EC  & 15 & $[20, 50]$ & $\times$ \\ 
		Packing problems  & PP  & 18 & $[8, 48]$ & $\times$ \\
		Airfoil design  & AD  & 5 & $[3, 34]$ & \checkmark \\ 
		Flow fitting  & FF  & 8 & $[6, 16]$ & \checkmark \\ 
		Hydraulic network optimization  & HN  & 8 & $[8, 58]$ & \checkmark \\ 
		Structural frame design problems  & SFD  & 13 & $[4, 24]$ & \checkmark \\ 
		Ill-posed problems  & IPP  & 24 & $[5, 50]$ & $\times$ \\
		\hline
	\end{tabular}
	\label{tab:IndagoBenchFuns}
\end{table}

The test suite includes a subset of the CEC 2014 test suite (namely $D \in \{10, 20, 50\}$). CEC test suites have, despite having some drawbacks (as previously explained), been continuously used and are now de facto standard for optimization research. Including CEC 2014 in IndagoBench25 widens the scope of the test suite by significantly increasing the test function count and more importantly by combining the purposely designed syntethic CEC functions with the realistic engineering-based problems.

The distribution of IndagoBench25 functions in terms of dimensionality $D$, multimodality $M$ (explained in Section \ref{sec:analysis}) and difficulty (average $\mathbb{G}$) is given in Figure \ref{fig:functions_scatter}. All test functions are explained in detail in the supplementary material. 

\begin{figure}[h]
	\centering
	\includegraphics[width=1\linewidth, trim={3mm 3mm 10mm 3mm}, clip]{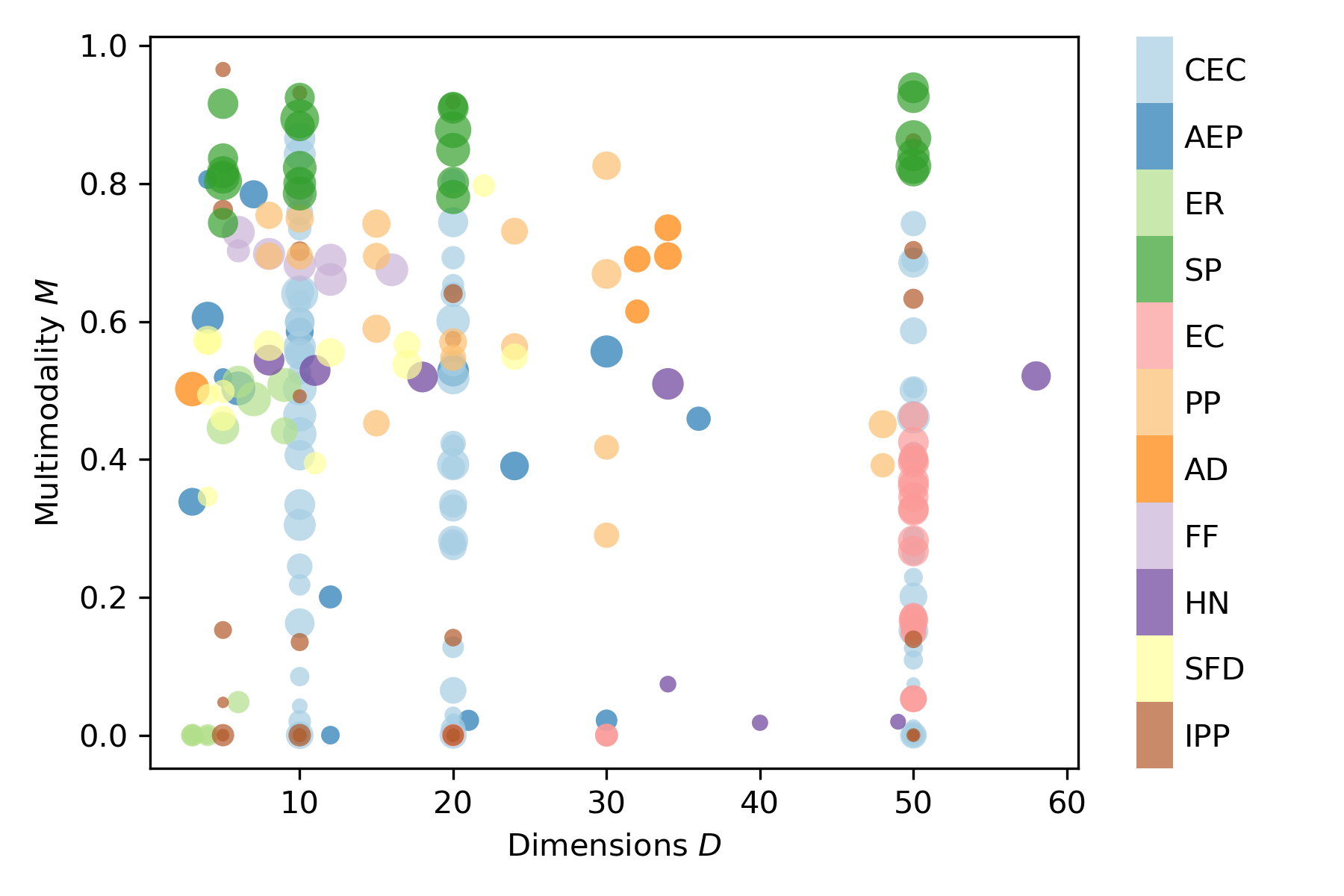}
	\caption{Distribution of the IndagoBench25 test functions. Function multimodality $M$ is assessed as per \eqref{eq:multimodality}. Point size is defined by average $\mathbb{G}$ across all optimization methods (larger circle means harder problem, i.e. lower $\mathbb{G}$).}
	\label{fig:functions_scatter}
\end{figure}

The 20 chosen optimization methods were tested on all of the IndagoBench25 test functions. In order to properly apply the $\mathbb{G}(f)$ metric, both RS (of which the median result was used as the $f^\circ$ reference point) and all other optimization methods used the exact same number of evaluations. 

Furthermore, for each optimization method on each test function, the statistical stability of the obtained $\mathbb{G}$ scores was ensured by observing the convergence of the median $\mathbb{G}$ scores for fitness achieved after 100\% of evaluations $f_{100\%}$, over $r$ runs of the optimization:
\begin{align*}
	\mathbb{G}_{med}^{(r)} = \operatorname{med}\left\{\mathbb{G}(f_{100\%}^{(1)}), \ldots, \mathbb{G}(f_{100\%}^{(r)})\right\}.
\end{align*}
To estimate the convergence and stability $\mathbb{G}_{med}^{(r)}$ as the total number of runs increases, its variability over the 50 last runs was measured as 
\begin{align*}
	\varepsilon = \max\left\{\mathbb{G}_{med}^{(r-50)},\ldots, \mathbb{G}_{med}^{(r)}\right\} -\min\left\{\mathbb{G}_{med}^{(r-50)},\ldots,\mathbb{G}_{med}^{(r)}\right\} 
\end{align*}
and the number of optimization runs $r$ was increased (in batches of 10) until the criteria $\varepsilon \leq 0.01$ and $r \geq 100$ are met. When the convergence defined in this way is ensured, the relevant metric $\mathbb{G}_{\mathrm{med}}^{(r)}$, i.e. the median value across all performed optimization runs remains within a strict tolerance. 
A similar procedure was employed for producing the random sampling median value $f^+$, with the same $\varepsilon$ criteria applied on at least 2000 evaluation trials. 

The median performance across independent optimization runs serves as an indicator of optimization quality at different search stages or under restricted computational budgets. We define $\mathbb{G}_{10\%}$, $\mathbb{G}_{50\%}$, and $\mathbb{G}_{100\%}$ based on the achieved results after 10\%, 50\%, and 100\% of the total function evaluations. Similarly, we define the corresponding relative metrics $\overline{\mathbb{G}}_{10\%}$, $\overline{\mathbb{G}}_{50\%}$, and $\overline{\mathbb{G}}_{100\%}$, which utilize the RS reference values $f^\circ$ obtained at the same stages of the optimization process. Note that all of these metrics can be calculated from the results of an optimization using maximum available evaluations for a given problem. In the following discussion, $\mathbb{G}$ is, for notation convenience, taken to be equivalent to $\mathbb{G}_{100\%}$.
 
In statistical processing of multiple optimization runs it is interesting to observe Empirical (or sample) Cumulative Distribution Function (ECDF) of obtained $\mathbb{G}$ values. Its inverse, the Empirical (or sample) Quantile Function (EQF), allows for an overview of performances across multiple optimization runs and it can be suitably visualized using percentile $\mathbb{G}$ plot as seen in Figure \ref{fig:example_convergence_plot}. We define EQF as $\widetilde{\mathbb{G}}(p)$ where $p \in [0, 1]$ is the percentile, i.e. the threshold below which $p$'s of the obtained $\mathbb{G}$'s fall. By definition, $\widetilde{\mathbb{G}}(0.5)$ is the median of sampled $\mathbb{G}$ values giving $\mathbb{G}_{100\%}$. Analyzing the EQF of $\mathbb{G}$ allows us to statistically evaluate the performance of repeated runs using the same optimization method, which is a viable strategy when sufficient computational resources are available. We define the repeating weighted metric $\mathbb{G}_{RW}$ that represents an anticipated $\mathbb{G}$ after 10 repeated optimizations:
\begin{equation*}
	\mathbb{G}_{RW} = \frac{\sum_{i=1}^{10}(\widetilde{\mathbb{G}}(0.5 ^ {1 / i}) \cdot w_i)}
	{\sum_{i=1}^{10}(w_i)} ~,
\end{equation*}
with suitably decreasing weights for each forthcoming run, defined as $w_i = 1 / i$. Since the EQF $\widetilde{\mathbb{G}}$ is monotonically increasing, the repeated weighted $\mathbb{G}_{RW}$ is always equal or greater than the median $\mathbb{G}_{100\%}$ corresponding to a single optimization run.


\section{Benchmark results for a selection of engineering problems}

A selection of characteristic IndagoBench25 test problems is briefly described in the following section, with some interesting observations on the optimization results. The problems illustrated below are all explained in detail in the supplementary material (see Section \ref{sec:supplementary}). Since the engineering-type and especially simulation-based test functions are particularly interesting for assessing real-world performance of the tested optimization methods, an emphasis is given here accordingly. 

\begin{figure}[h!]
	\centering
	\includegraphics[width=1\linewidth, trim={0mm 0mm 0mm 10mm}, clip]{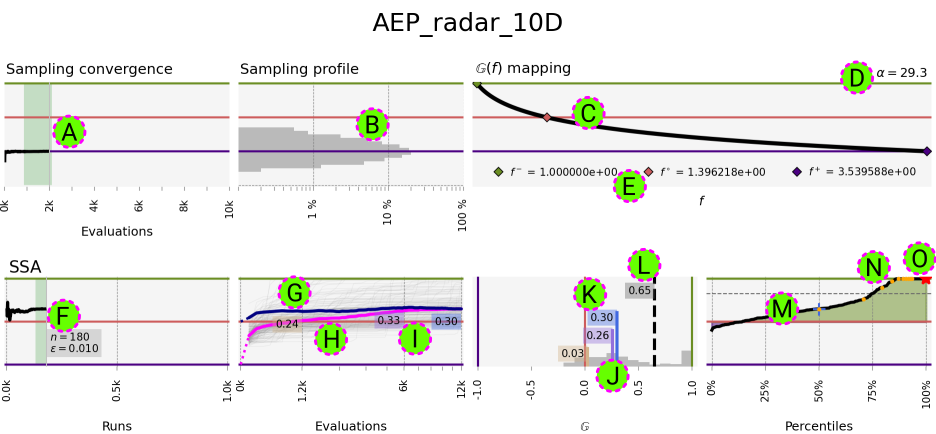}
	\caption{Subplots in upper panel show function information, the distribution \textbf{(B)} of random sampling and the convergence of its median value \textbf{(A)}. The reference points ($f^+$, $f^\circ$ and $f^-$) \textbf{(E)} are used to determine $\mathbb{G}(f)$ mapping function \textbf{(C)} along with its parameter $\alpha$ \textbf{(D)}. Lower panel display a benchmark for a selected optimization method (SSA). Statistical convergence of median $\mathbb{G}$ over multifold optimization runs  is tracked in \textbf{(F)}. Convergence of median $\mathbb{G}$ over evaluations is shown with magenta curve \textbf{(H)}, while convergence of its relative value (when reference RS point $f^\circ$ for same number of evaluations is used) is shown with dark blue curve \textbf{(G)}. The values of relative $\mathbb{G}$ for 10\%, 50\% and 100\% of evaluations are shown in labels \textbf{(I)}. The distribution of $\mathbb{G}_{100\%}$ is shown with gray bins \textbf{(J)}. Vertical lines and labels \textbf{(K)} indicate median values for $\mathbb{G}_{10\%}$, $\mathbb{G}_{50\%}$ and $\mathbb{G}_{100\%}$. Black dashed line and its label \textbf{(L)} exhibit value of repeating weighted $\mathbb{G}_{RW}$ which is obtained from sampling (orange) points \textbf{(N)} indicated in $\mathbb{G}_{100\%}$ percentile graph \textbf{(M)}. If best overall solution is found by the method, it is indicated by the red star in the  percentile graph \textbf{(O)}. All of these plots contain a $\mathbb{G}$ axis, indicated with purple, red and green lines specifying reference points of the $\mathbb{G}$ metric.}
	\label{fig:example_convergence_plot}
\end{figure}

To analyze optimization performances, a composite plot of convergence curves and statistical distribution of results is used. The explanation of such plot is given in Figure~\ref{fig:example_convergence_plot}. 

\subsection{Shortest path problem example}

The shortest path (SP) problem is formulated by representing the path as a polyline of equal-length segments defined by a vector of relative angles, ensuring the path connects fixed start and end points efficiently. The method includes scaling and rotating the initial unit path to match the target endpoints while allowing flexibility of path shape. Obstacle avoidance is handled using a penalty-based fitness function that adds costs for path segments intersecting circular obstacles, thereby enforcing collision-free shortest paths. A variety of obstacle arrangements and numbers of path segments are considered across 24 analyzed problems, allowing for a clear and intuitive visual interpretation of the problem’s multimodality.

\begin{figure}[h!]
	\centering
	\includegraphics[width=1\linewidth, trim={0mm 0mm 0mm 10mm}, clip]{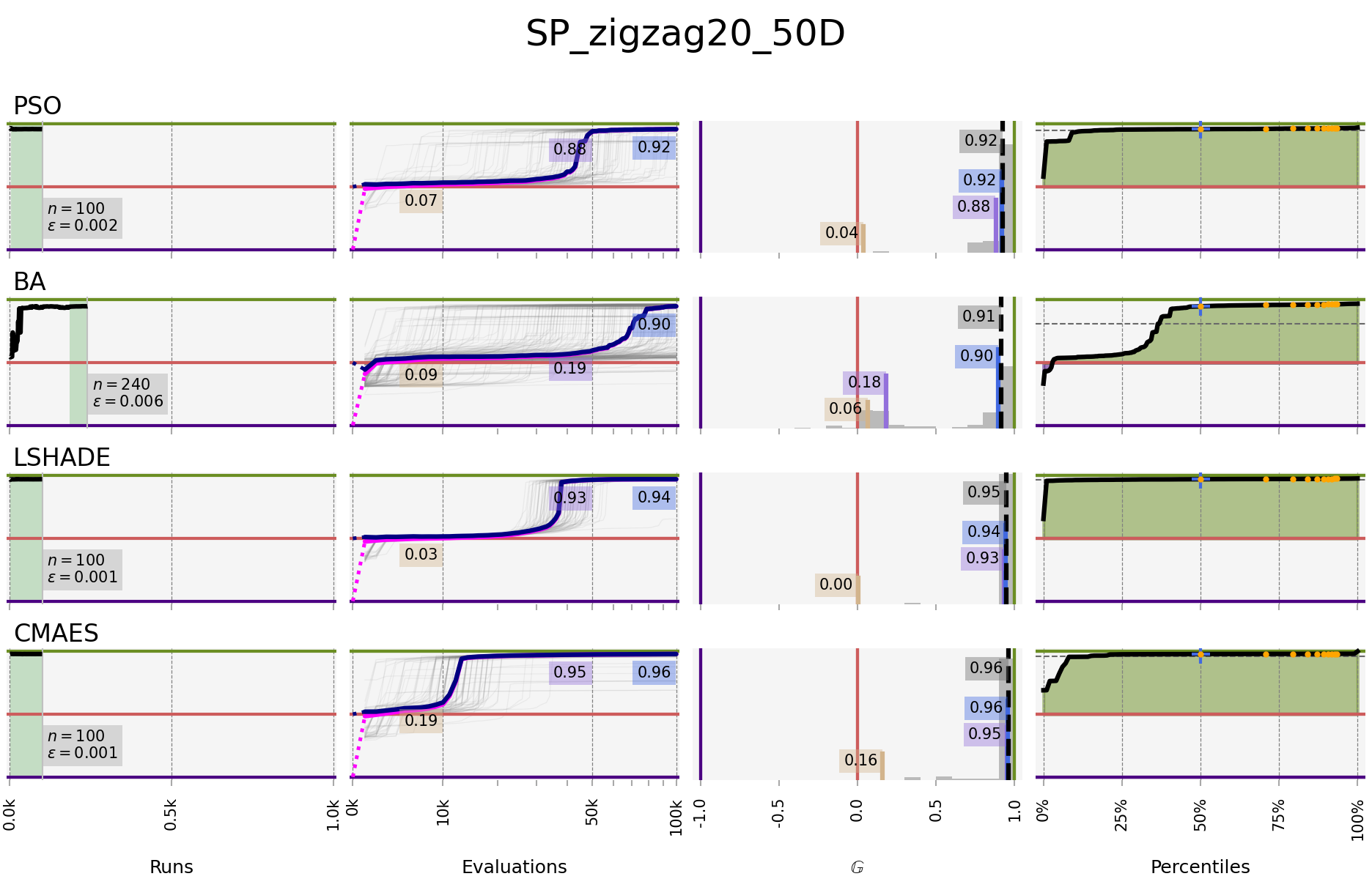}	
	
	\vspace{3mm}
	\includegraphics[width=1\linewidth]{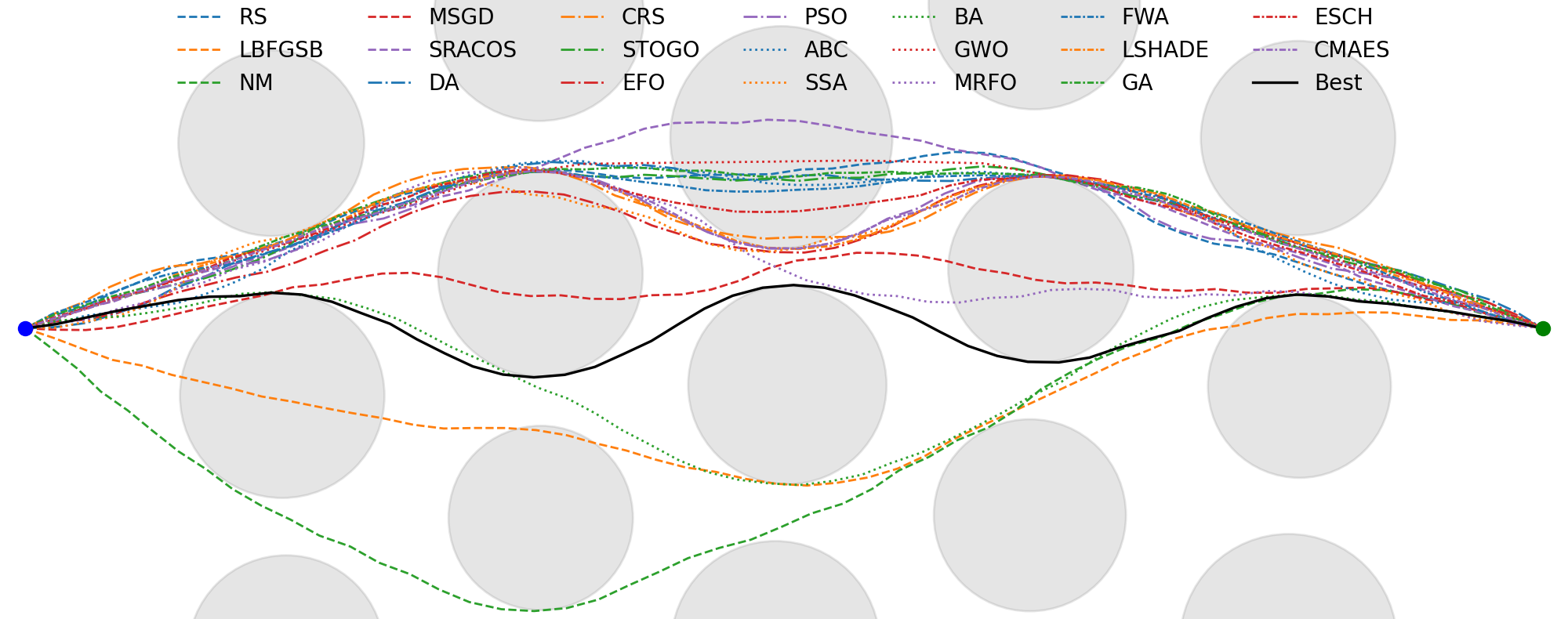}
	\caption{The convergence of selected optimization methods for shortest path problem \texttt{SP\_zigzag20\_50D}. Out of four median solutions, although achieving very high $\mathbb{G}$, none converged to global optima, as visually represented with distinctive difference in depicted path patterns. Switching between local optima can be observed in steep jumps in the shown convergence plots.}
	\label{fig:SPexample}
\end{figure}

Figure~\ref{fig:SPexample} shows achieved median solutions for all methods and overall best solution for the \texttt{SP\_zigzag20\_50D} problem case. The diversity of methods' performances can be easily visually recognized, with achievement of locally optimal paths and obstacles collision constraint breaches. Avoiding obstacle collision is manifested in rapid increases in $\mathbb{G}$ during the optimization. Interestingly, many of the optimization methods fail to deliver a feasible solution i.e. non-colliding path.

Across SP problems, LSHADE and CMAES consistently exhibited superior performance, with CRS also performing strongly in certain cases. Interestingly, NM often achieved very competitive results with only a few repeated optimization runs.

\subsection{Ergodic coverage problem example}

The ergodic coverage (EC) problem aims to design a trajectory that distributes a robot’s time across a domain proportionally to a prescribed goal density, ensuring thorough and weighted exploration. Coverage quality is evaluated by using either $L_1$, $L_2$ or spectral metrics. The trajectory is parameterized as in the SP problems but without a fixed endpoint, offering an intriguing heterogeneous-variable problem suitable for optimization benchmarking.

\begin{figure}[h!]
	\centering
	\includegraphics[width=1\linewidth, trim={0mm 0mm 0mm 10mm}, clip]{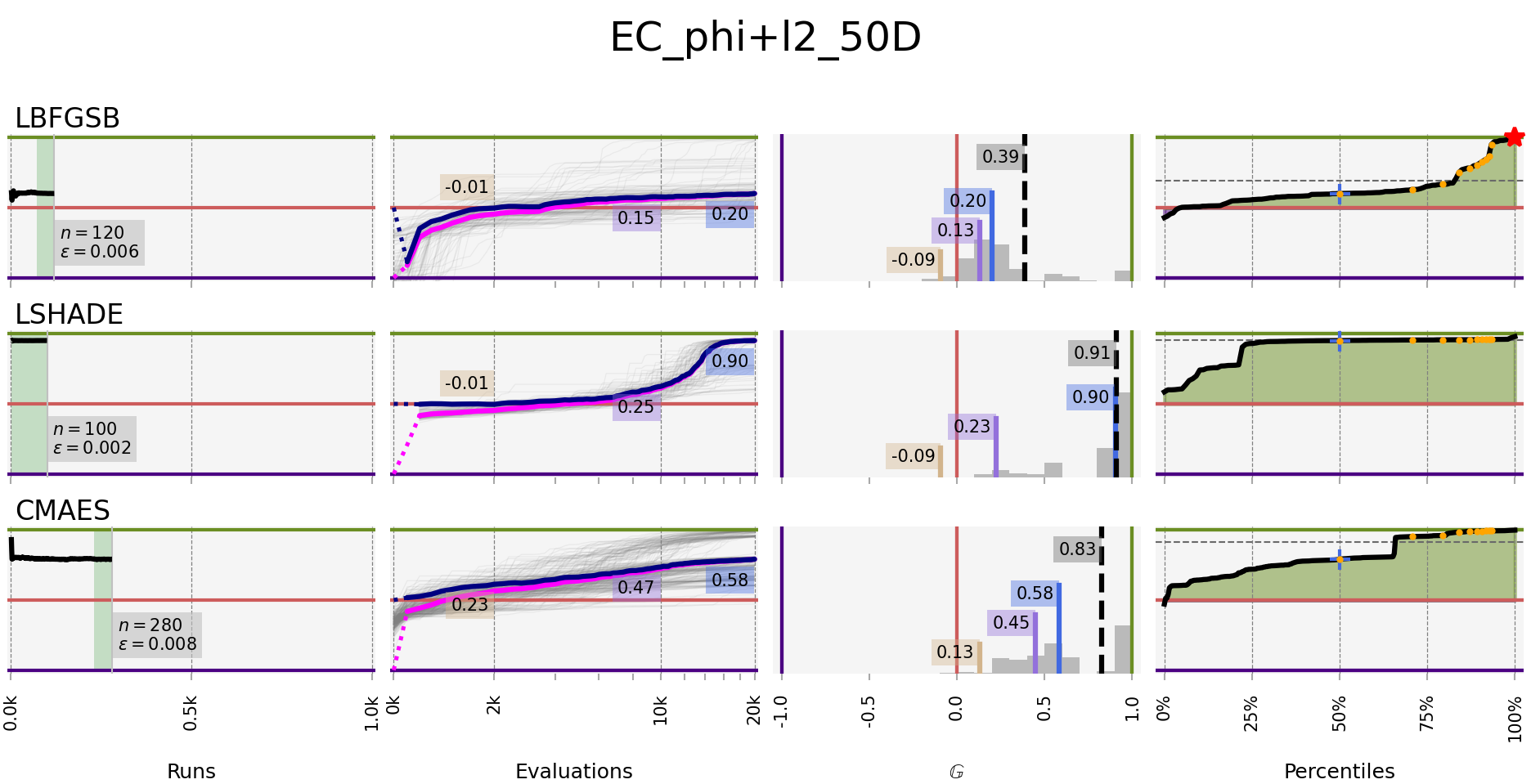}	
	
	\vspace{3mm}
	\includegraphics[height=0.338\linewidth]{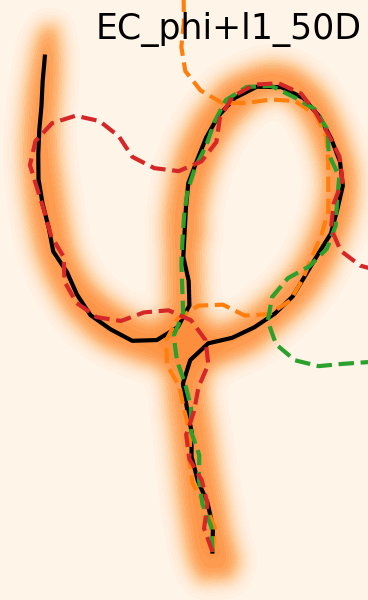}
	\includegraphics[height=0.338\linewidth]{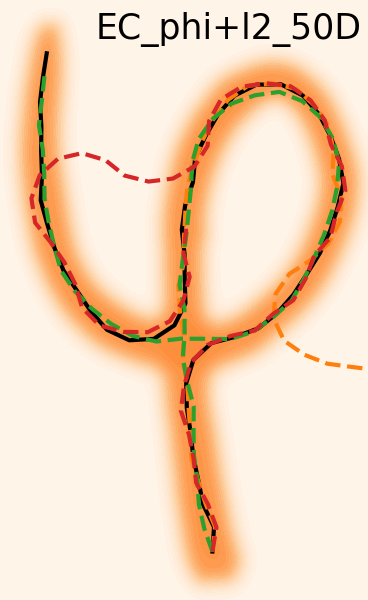}
	\includegraphics[height=0.338\linewidth]{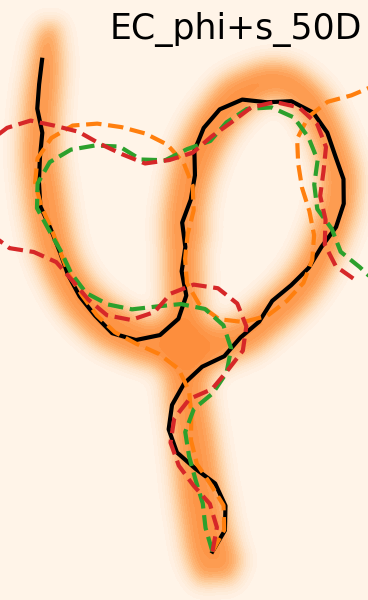}
	\includegraphics[height=0.338\linewidth]{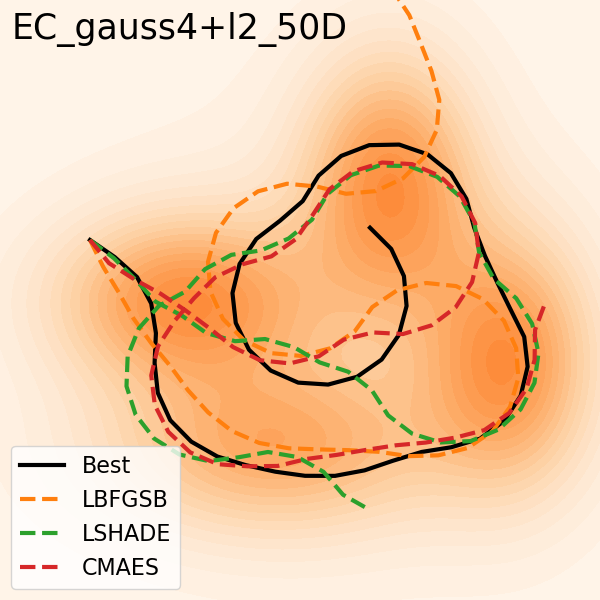}
	\caption{The convergence of selected optimization methods for ergodic problem \texttt{EC\_phi+l2\_50D} shows interesting potential of repeated runs of LBFGSB method. The three bottom-left images clearly depict the difference in $\mathbb{G}$ score as related to the different formulations of ergodic coverage problem and shows advantage when using \texttt{l2} formulation in comparison to \texttt{l1} or \texttt{s}. The bottom-right plot reveals that  unfavorable results are achieved on \texttt{EC\_gauss4+l2\_50D} problem regardless of the used optimization method.}
	\label{fig:ECexample}
\end{figure}

Convergence plots for selected methods on the \texttt{EC\_phi\_50D} problem are shown in Figure~\ref{fig:ECexample}. The problem multimodality manifests in dispersed $\mathbb{G}_{100\%}$ distributions.

In general, the EC problems proved to be highly challenging, with all optimization methods exhibiting mixed and inconsistent performance. Some cases appeared practically unsolvable under the benchmark configurations used. However, repeated optimization runs often resulted in significant improvements across most EC problems for nearly all methods. This behavior is particularly interesting for stochastic population-based algorithms, which are typically considered more robust than methods focused on local search.

\subsection{Packing problem example}

\begin{figure}[h!]
	\centering
	\includegraphics[width=1\linewidth, trim={0mm 0mm 0mm 10mm}, clip]{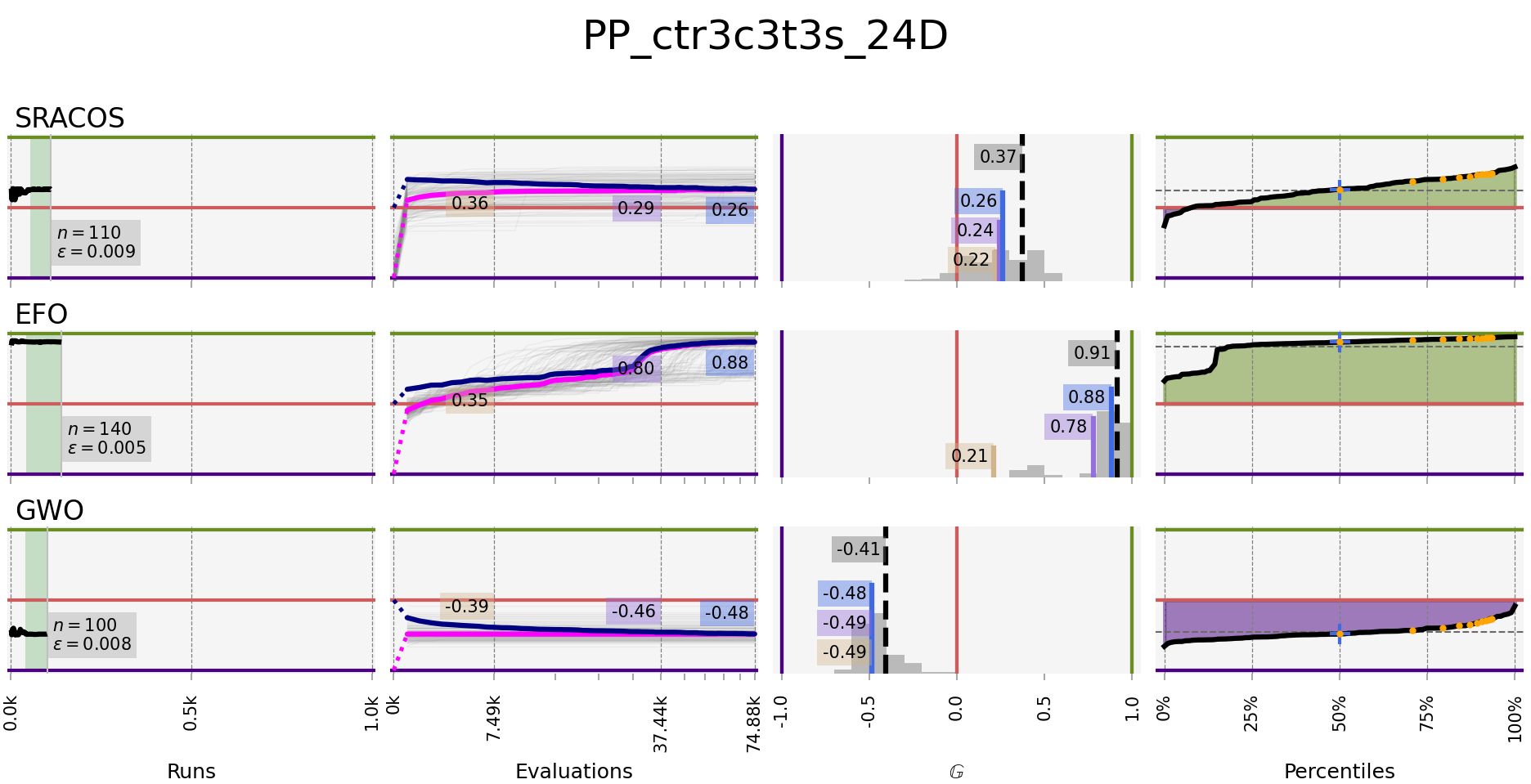}
	\\[-2mm]
	\valign{#\cr
		\hsize=0.5\linewidth
		\subfloat{\includegraphics[width=\hsize, trim={22mm 20mm 10mm 17mm}, clip]{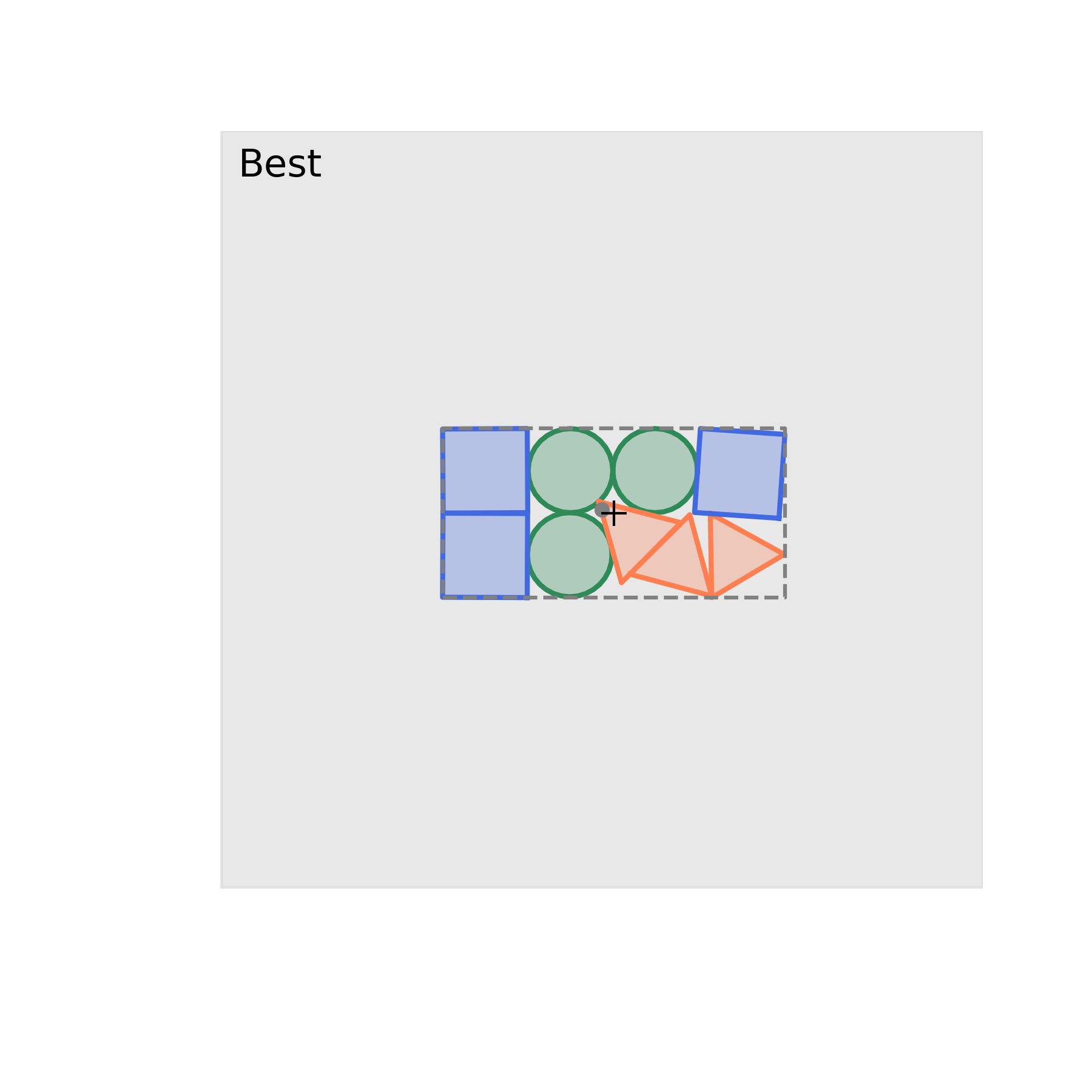}}\\[-5mm]\vfill
		\subfloat{\includegraphics[width=\hsize, trim={22mm 20mm 10mm 17mm}, clip]{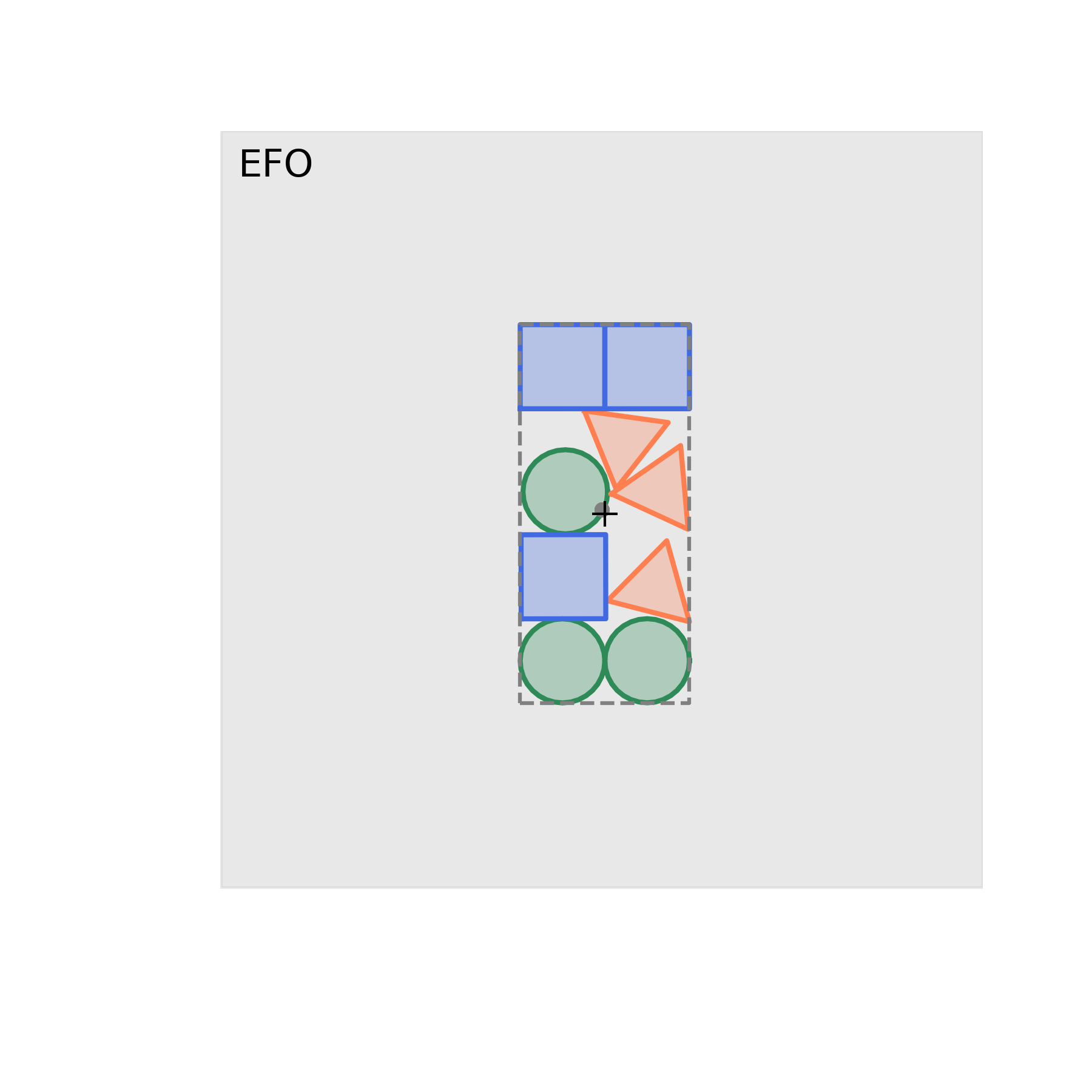}}\\[-3mm]\cr
		\hsize=0.5\linewidth
		\subfloat{\includegraphics[width=\hsize, trim={22mm 20mm 10mm 17mm}, clip]{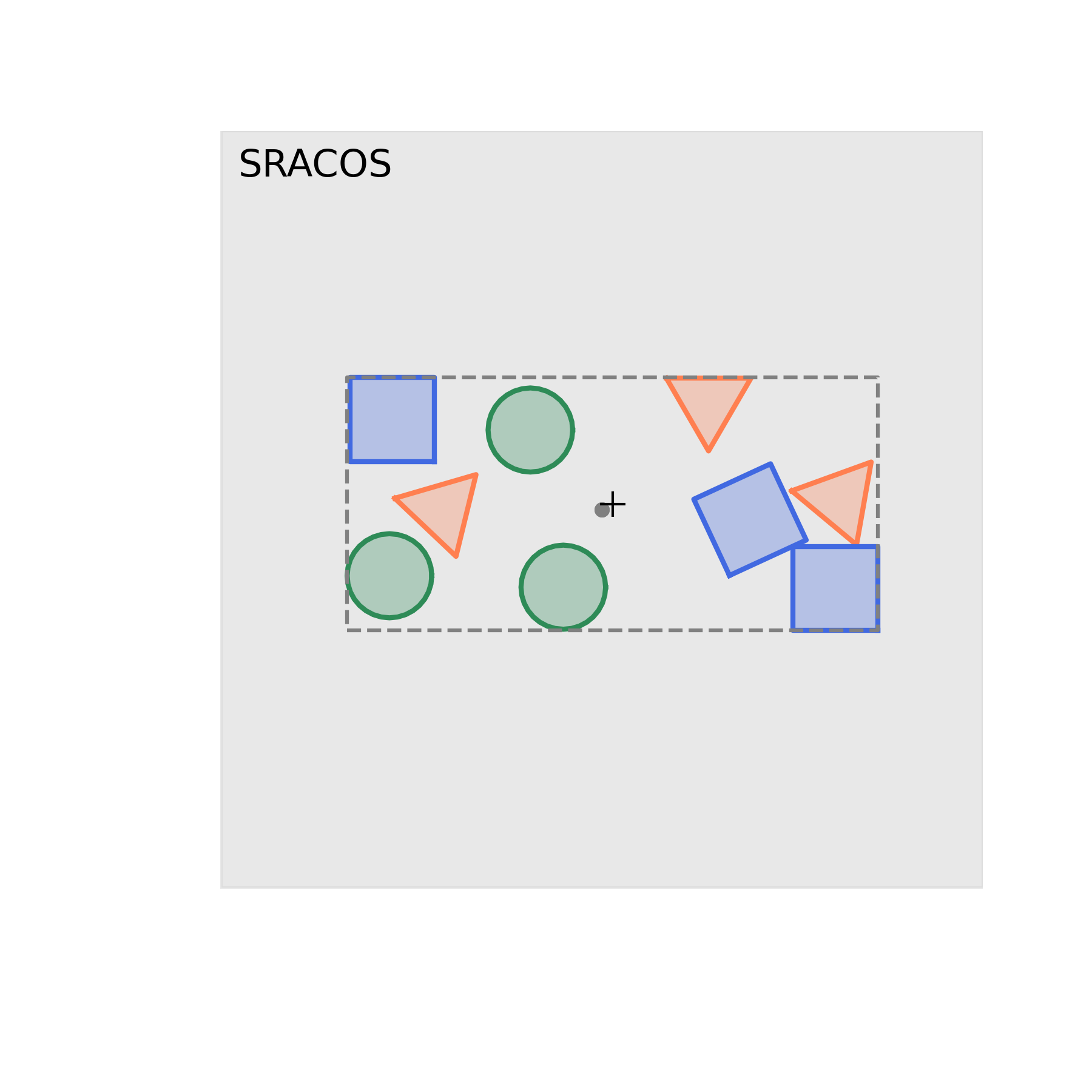}}\\[-5mm]\vfill
		\subfloat{\includegraphics[width=\hsize, trim={22mm 20mm 10mm 17mm}, clip]{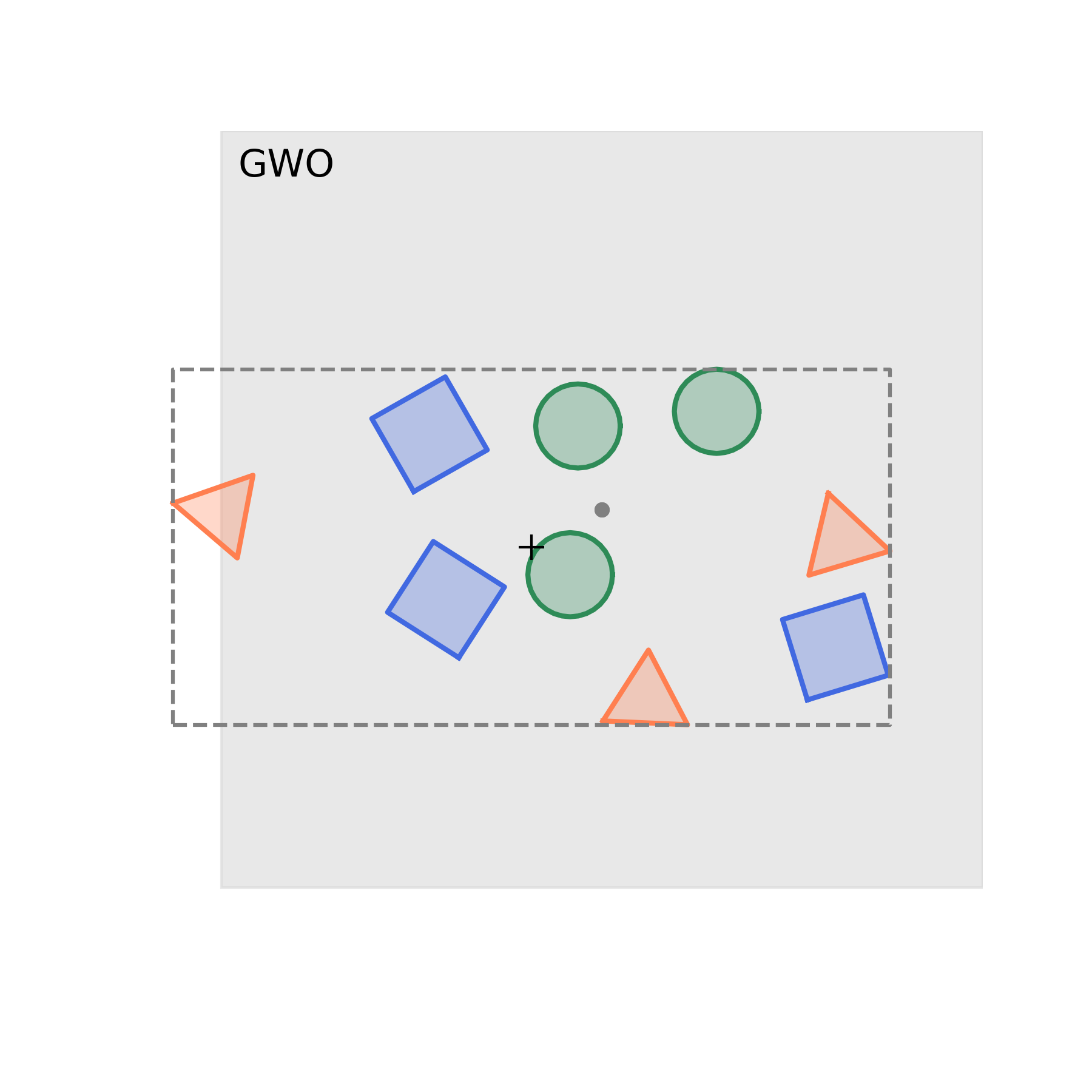}}\\[-3mm]\cr
	}
	\caption{Convergence plots and best results for the \texttt{PP\_ctr3c3t3s\_24D} function, for a selection of the tested optimization methods. Grey area represents the sheet (domain), dashed rectangular is the bounding box of the shapes, black plus sign is the center of the bounding box, and grey dot is the center of the sheet. It can be observed that, while some methods (EFO) solved the problem reasonably well, others (SRACOS) performed rather poorly, and some were even catastrophically inefficient (GWO performing worse than RS). Comparing the overall best found solution with the EFO result, it is obvious that the problem is highly multimodal.}
	\label{fig:PPexample}
\end{figure}

Packing problems (PP) are dealing with grouping objects together into containers, without overlap. As an example from the PP problem group, an excerpt of the results for the \texttt{PP\_ctr3c3t3s\_24D} function is shown in Figure \ref{fig:PPexample}. The optimization task for this function is to tightly group a collection of 9 geometric shapes (3 circles, 3 triangles, and 3 squares) at the center of the sheet. 

PP cases are hybrids of a smooth local problem (positioning the shapes in the center of the domain) and a combinatorial problem (arranging the shapes into one of the many optimal arrangements), and as such have a huge (albeit finite) number of solutions, some of which are interchangeable.
The PP optimization have shown to be rather difficult, as most methods failed to find a good solution for them. Moreover, for some of these problems none of the methods confidently succeed in solving them. When a PP problem is solved, it is mostly by SRACOS, CRS, LSHADE or CMAES, however without any obvious consistency in their specific performance across the entire PP problem class.

\subsection{Flow fitting problem example}

The flow fitting (FF) problems use CFD simulations to reconstruct sea-surface velocity fields from sparse buoy measurements by optimizing boundary conditions of a partial differential equation. Boundary pressures and tangential velocities are parameterized with cubic splines, producing a continuous optimization problem with heterogeneous variables. The objective function combines the average velocity mismatch between simulated and measured data with penalty terms enforcing simulation stability and physical consistency.

\begin{figure}[h!]
	\centering
	\includegraphics[width=1\linewidth, trim={0mm 0mm 0mm 10mm}, clip]{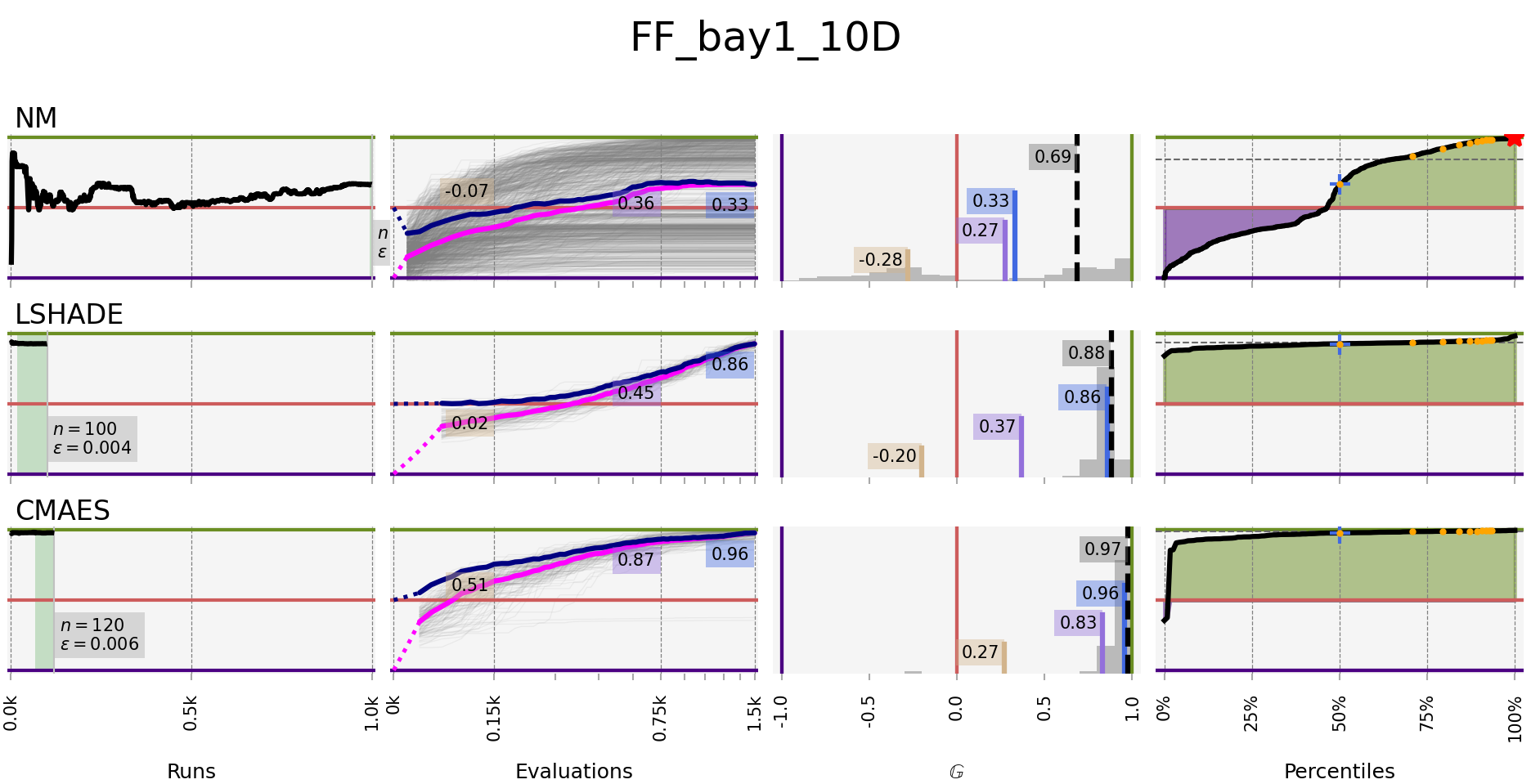}	
	
	\vspace{3mm}
	\includegraphics[height=0.49\linewidth]{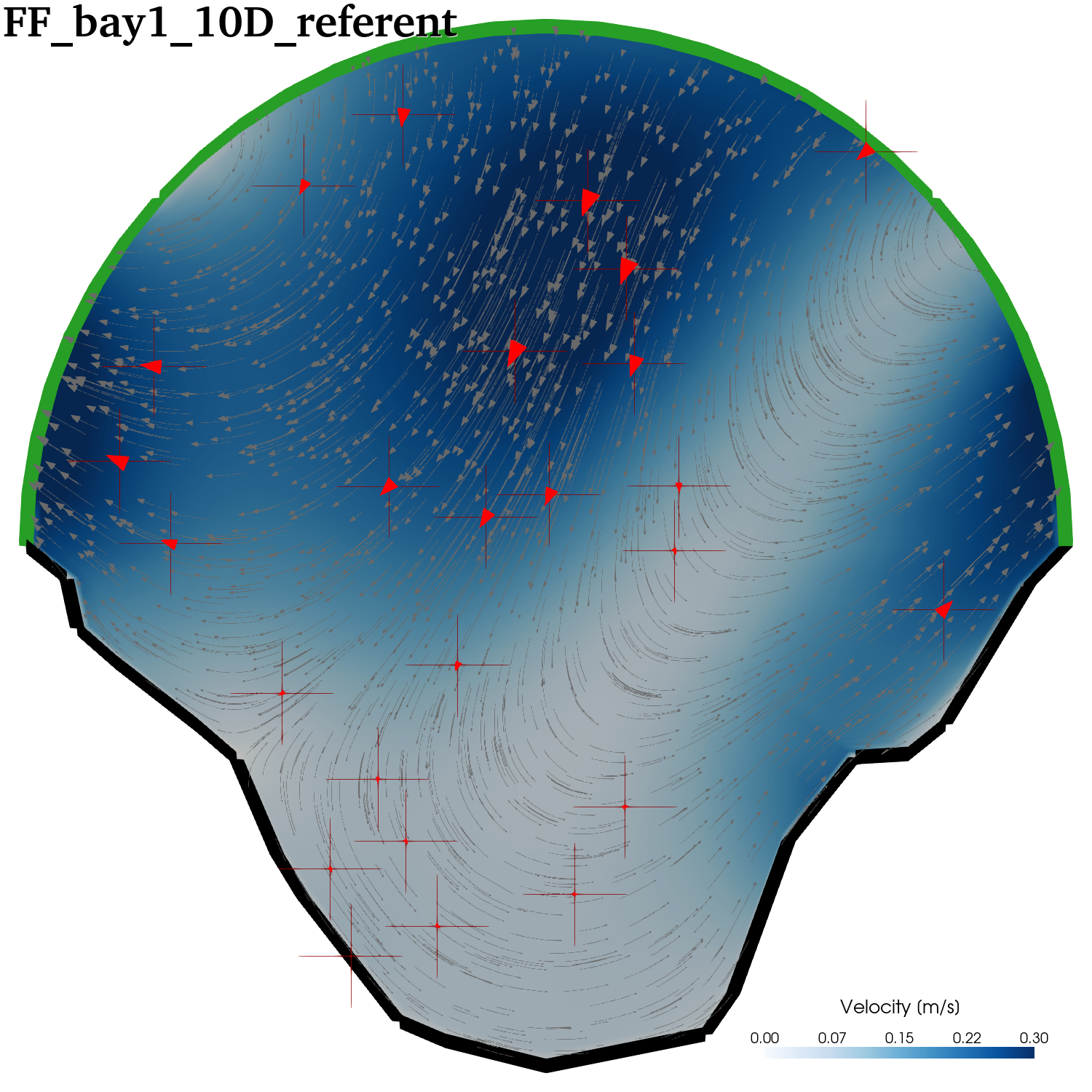}
	\includegraphics[height=0.49\linewidth]{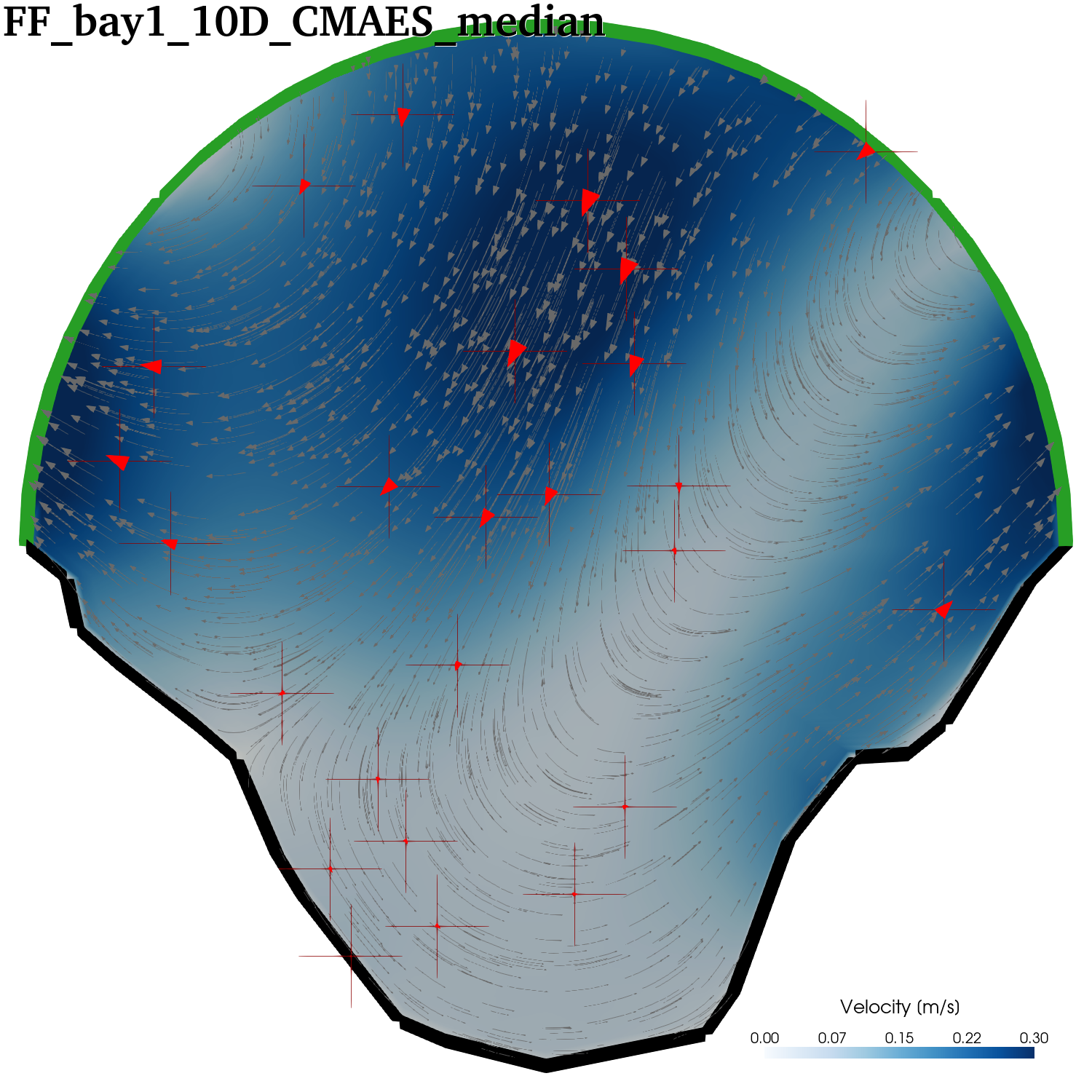}
	\caption{The convergence of selected optimization methods for the \texttt{FF\_bay1\_10D} problem shows diversity of method performances. The bottom images show the target flow field (left) and the CMAES-fitted flow field (right), with the red arrows representing reference velocities. Although with highly unstable median convergence, the best overall solution is found by NM (probably due to its exploitation of the high number of runs). Both LSHADE and CMAES are very robust and deliver consistent performance, though with noticeably different convergence trends. Comparison of the reference flow and median CMAES solution confirms high $\mathbb{G}_{100\%}=0.97$ score.}
	\label{fig:FFexample}
\end{figure}

The convergence behavior of selected optimization methods for the flow fitting problem \texttt{FF\_bay1\_10D}, shown in Figure~\ref{fig:FFexample},  demonstrates significant diversity in algorithmic performance. Since the FF problems rely on computationally intensive CFD simulations, the total number of evaluations was intentionally reduced to ensure feasible runtime; nevertheless, this reduction still proved sufficient for achieving accurate and representative results. In this test-case, the median convergence curve for NM exhibits considerable instability, which highlights potential shortcomings of using this metric for convergence control. CMAES, LSHADE, EFO, CRS, and to some extent GA, showed exceptional robustness, consistently achieving acceptable solutions in nearly every optimization run. Comparable performance was observed across other FF cases, though a few remained particularly challenging to solve.

\subsection{Hydraulic network optimization problem example}

The hydraulic network (HN) optimization problem focuses on determining optimal pipe diameters within a water distribution network, so as to minimize material usage while maintaining required pressures at consumer nodes. The network hydraulics are modeled using the EPANET solver, which computes flows and pressures under nonlinear constraints of mass and energy conservation based on the Hazen–Williams head-loss equation. The optimization implicitly refines both the network sizing and layout, as pipes with near-zero diameters are effectively removed from the design. A penalty-based fitness function combines normalized pipe diameters and pressure constraint violations, ensuring feasible, efficient, and hydraulically consistent network configurations.

\begin{figure}[h!]
	\centering
	\includegraphics[width=1\linewidth, trim={0mm 0mm 0mm 10mm}, clip]{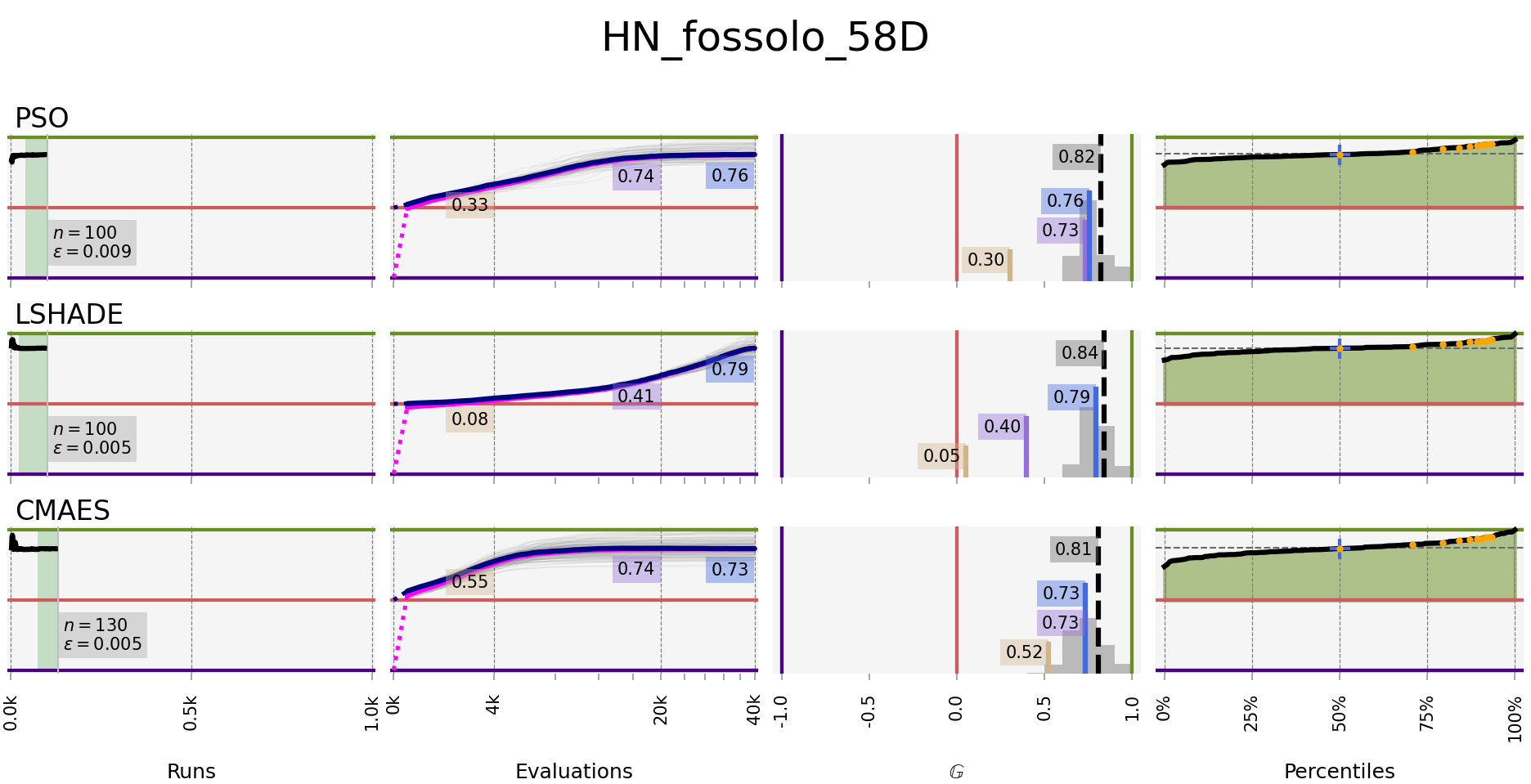}
	
	\vspace{3mm}
	\includegraphics[width=1\linewidth]{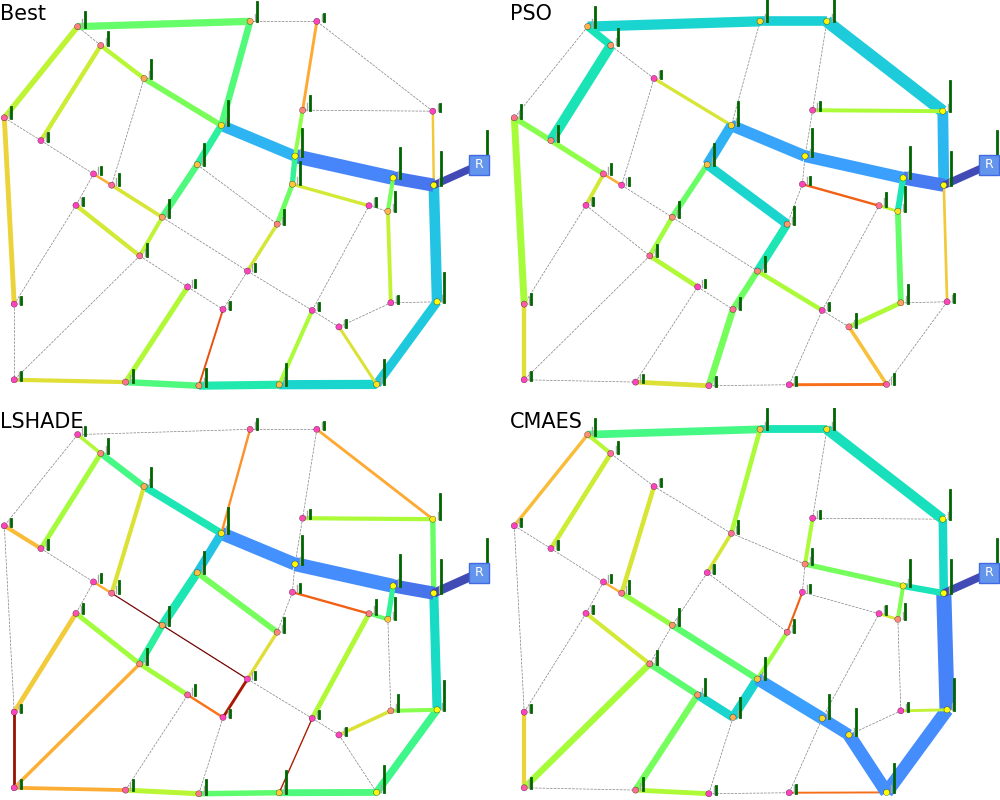}
	\caption{Distinguished convergence of LSHADE in comparison with other well-performing methods on the \texttt{HN\_fossolo\_58D} problem. Although some patterns of best network layout can be recognized in medial results, the achievement of optimal layout proves to be a hard problem. Even the best found solution does not seem to be optimal, given that a redesign in the top-left corner could produce a shorter network.}
	\label{fig:HNexample}
\end{figure}

Although it is formulated as a continuous pipe-sizing optimization task, its underlying nature is that of a network topology optimization problem. The \texttt{HN\_fossolo\_58D} problem shown in Figure~\ref{fig:HNexample} allows for multiple distinct network designs to achieve similar fitness values, making it inherently complex.

Although certain HN cases are more easily solved by several algorithms, LSHADE and EFO consistently demonstrate the most reliable performance across all HN problems.

\subsection{Structural frame design problem example}

The structural frame design (SFD) optimization problems deal with minimizing the total mass of planar beam-frame structures while satisfying engineering constraints on stress and displacement. Cross-section sizing and node positioning parameterize the structure design, while the evaluations are performed using CalculiX finite element simulations. Square beam cross-sections are used for simplicity, ensuring structural adequacy in a 2D setting. The constrained minimization task is reformulated into an unconstrained one via penalty functions, combining objectives for mass reduction with penalties for violating mechanical constraints. This allows for flexible definition of diverse structural optimization scenarios.

\begin{figure}[h!]
	\centering
	\includegraphics[width=1\linewidth, trim={0mm 0mm 0mm 10mm}, clip]{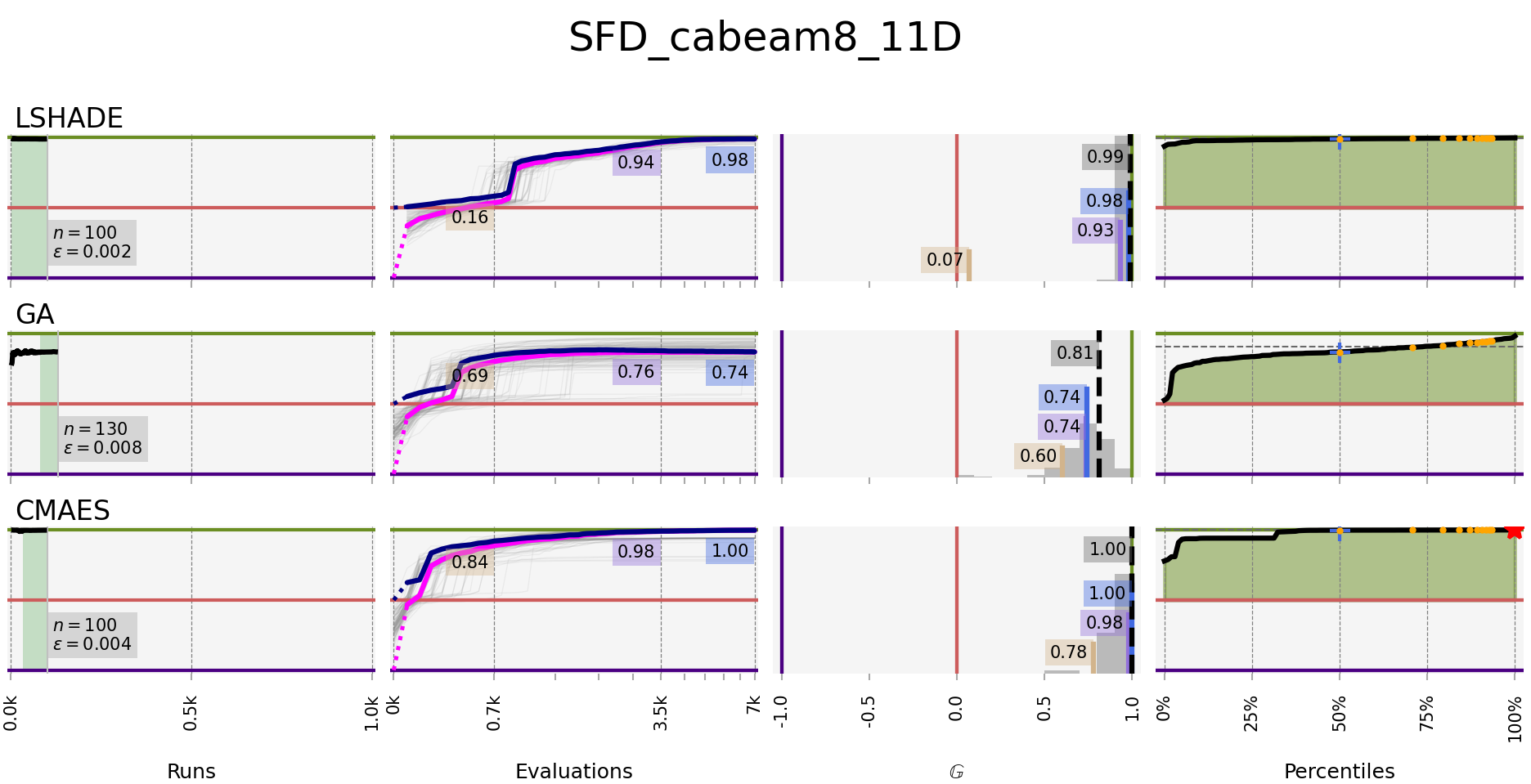}
	
	\vspace{3mm}
	\includegraphics[width=0.49\linewidth]{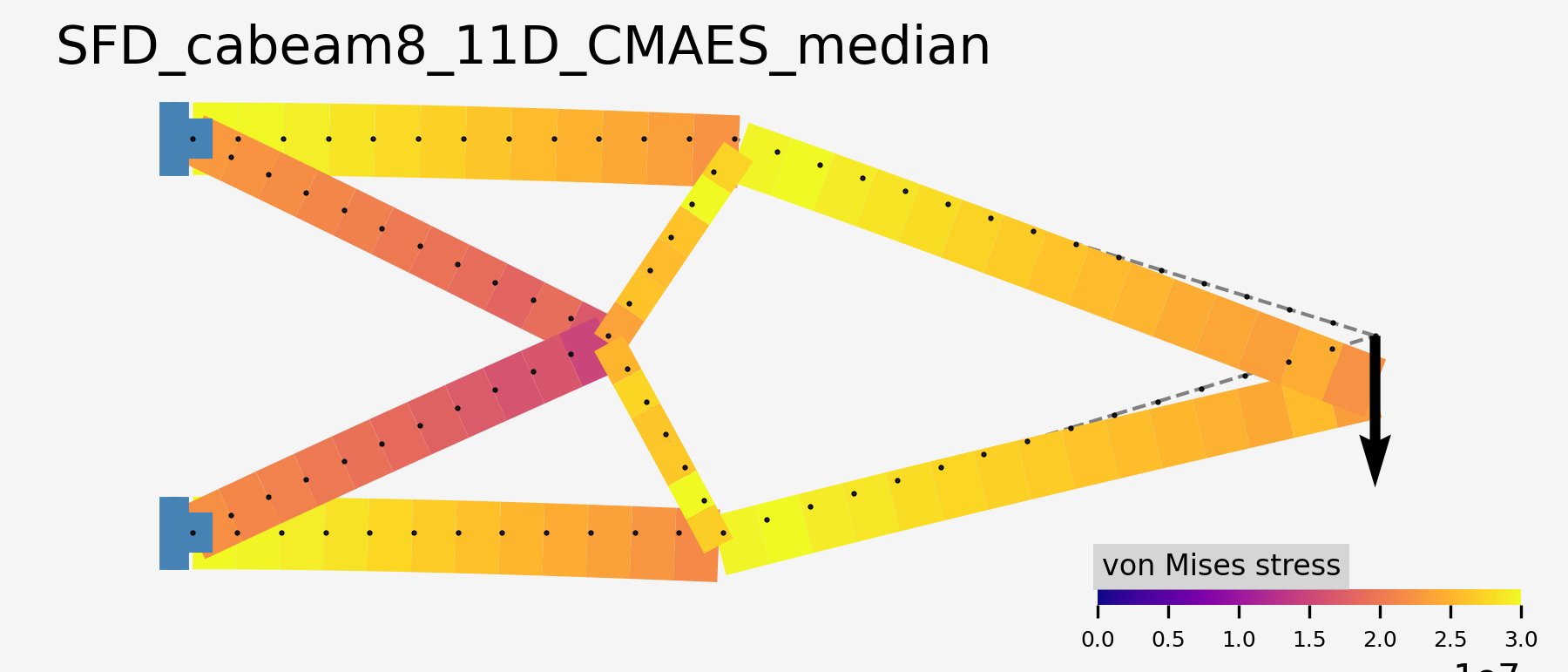} \hfill
	\includegraphics[width=0.49\linewidth]{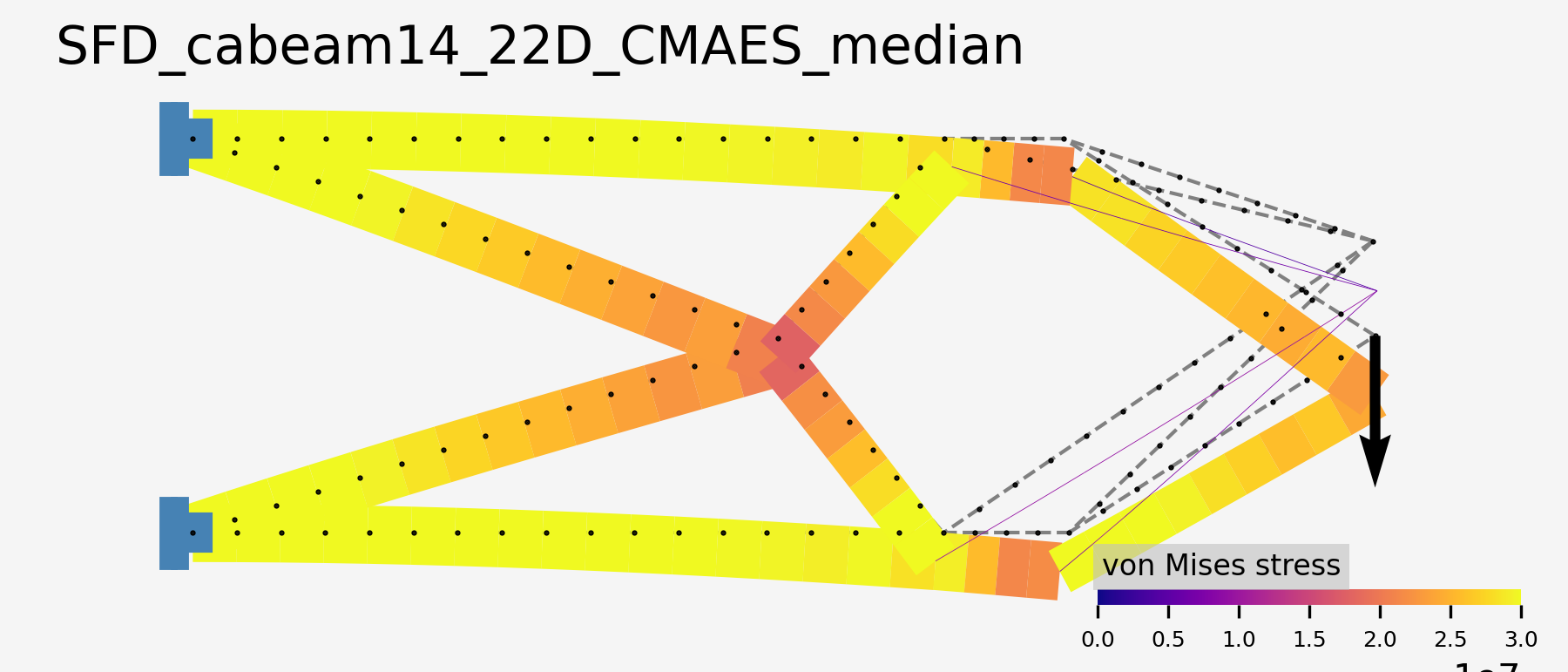}
		
	\vspace{1mm}
	\includegraphics[width=0.49\linewidth]{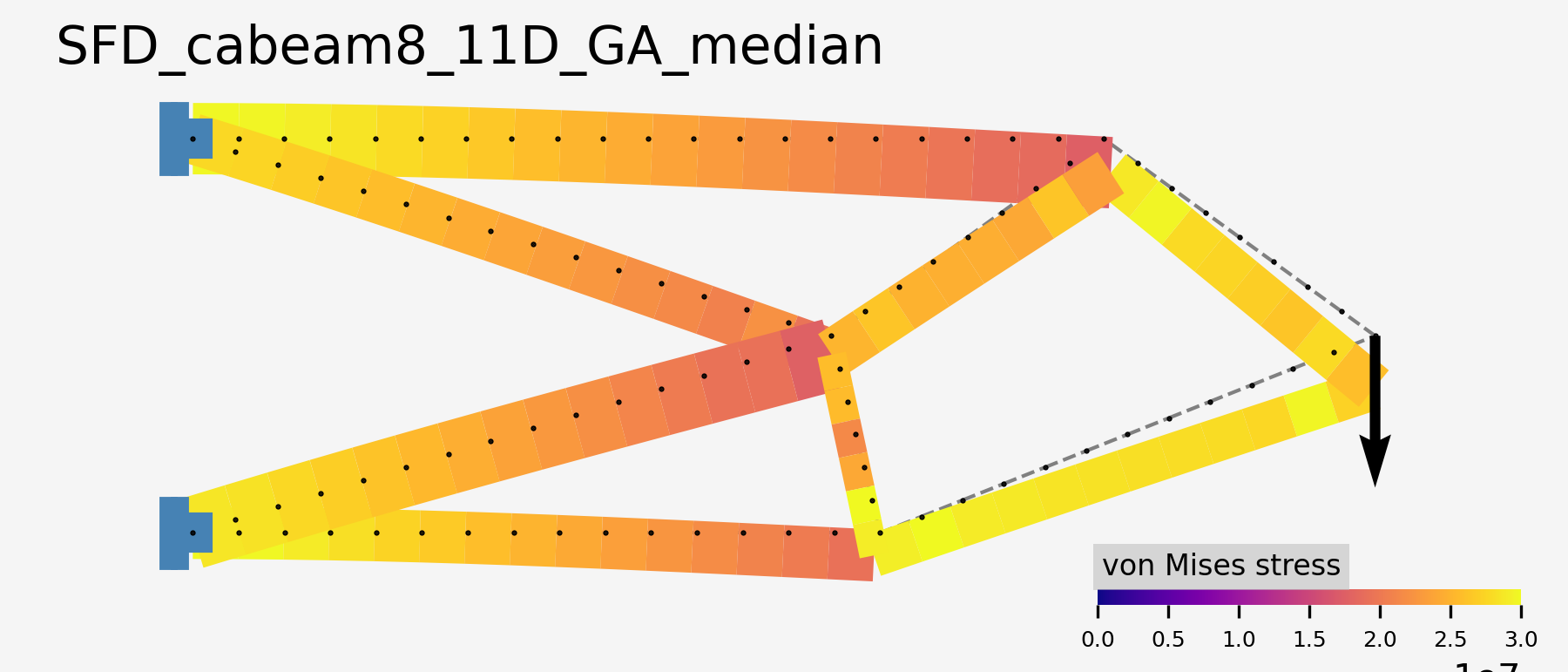} \hfill
	\includegraphics[width=0.49\linewidth]{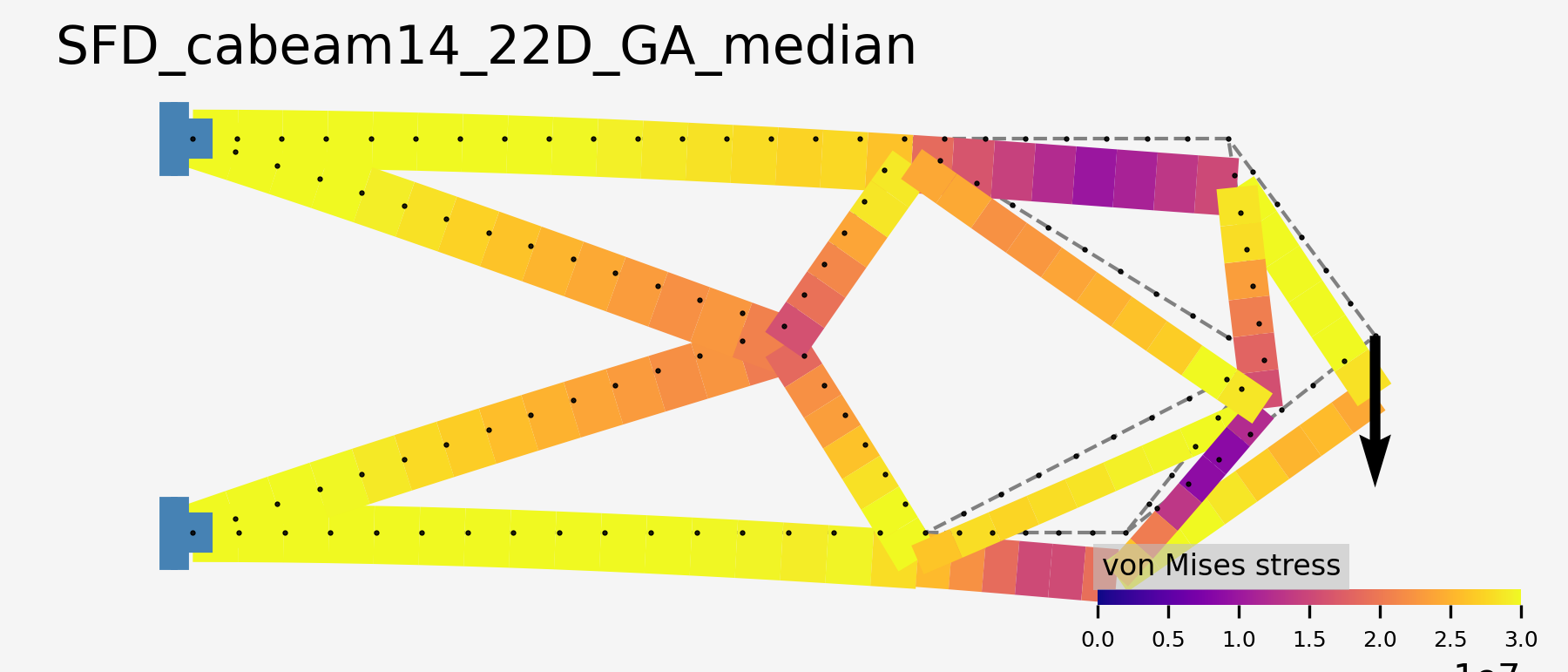}
	\caption{Convergence of best performing methods on \texttt{SFD\_cabeam8\_11D} problem shows different dynamics of constraint solving (note jumps in $\mathbb{G}$). However, faster constraint satisfaction does not guarantee better final results nor the robustness of a method (e.g. LSHADE vs. GA). Bottom panel displays selected \texttt{SFD\_cabeam8\_11D} and \texttt{SFD\_cabeam14\_22D} results, given in the left and right column respectively. The asymmetric designs are suboptimal, and it is interesting to observe that optimal solution for \texttt{SFD\_cabeam14\_22D} utilizes only 10 of total 14 beams and shows tendency to converge towards the \texttt{SFD\_cabeam8\_11D} design with 8 beams.}
	\label{fig:SFDexample}
\end{figure}

The best-performing methods convergence trends on the \texttt{SFD\_cabeam8\_11D} problem reveal distinct dynamics in constraint resolution (Figure~\ref{fig:SFDexample}). Yet, faster constraint resolving implies premature optimization which often harms the performance and robustness. The bottom panel illustrates selected results where asymmetric configurations appear suboptimal. 

Optimization methods solving SFD problems exhibited mixed performance, showing only partial success for both local and global search strategies. Overall, LSHADE and CMAES demonstrated the best and most consistent results, with EFO and CRS also performing well in several cases. In contrast, STOGO and MRFO struggled to achieve satisfactory outcomes in nearly all SFD problems.

\section{Method performance analysis}
\label{sec:analysis}

There are optimization methods which are more suited for solving specific types of optimization problems and then there are those that confidently outperform many others in the majority of use cases. In the following subsections we try to give a multifaceted and comprehensive analysis of optimization method performance. For brevity, $\mathbb{G}$ without a subscript is here used as a shorthand for $\mathbb{G}_{100\%}$.

\subsection{Overall method performance}

\begin{figure}[h!]
	\centering
	\includegraphics[width=1\linewidth, trim={3mm 3mm 7mm 5mm}, clip]{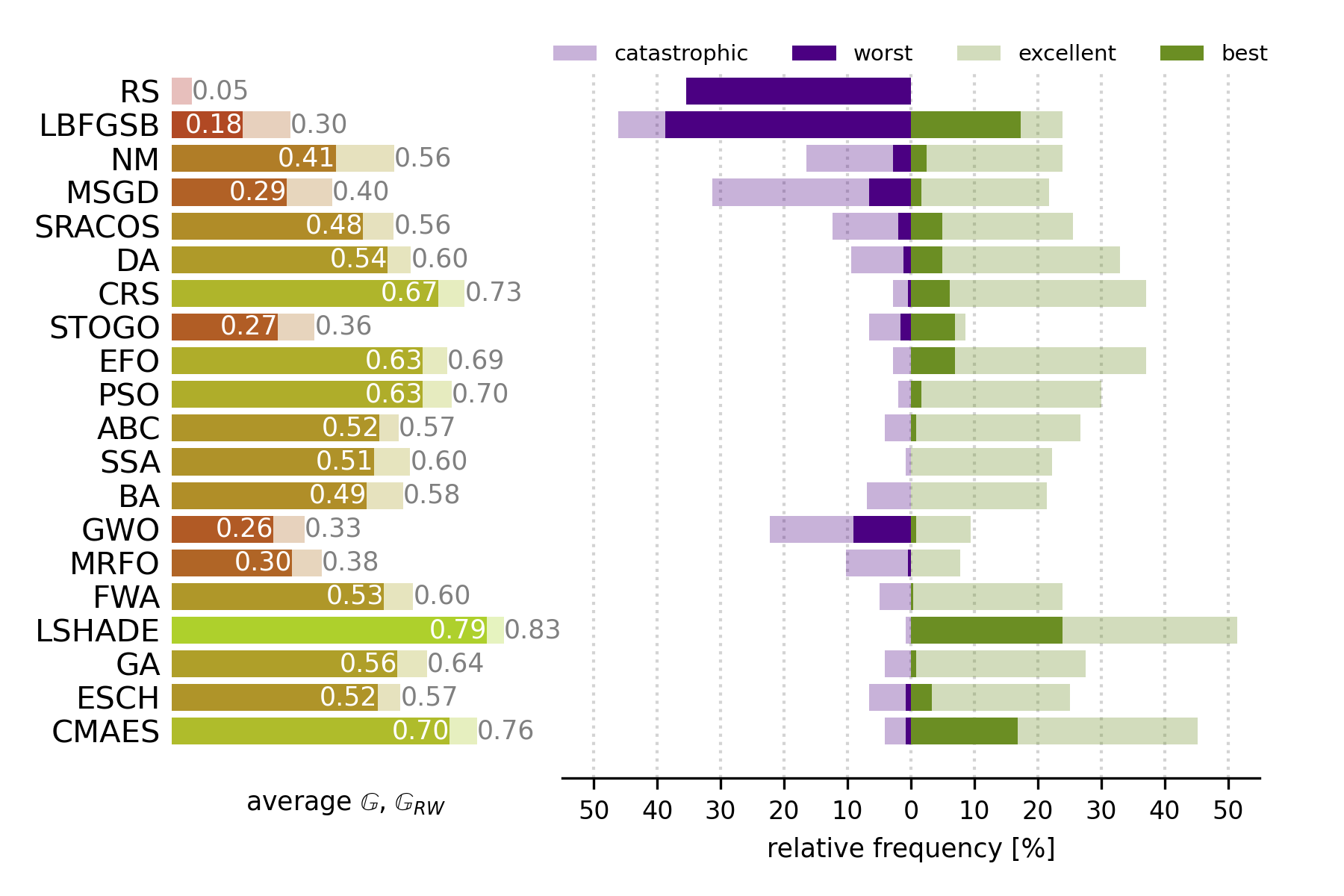}
	\caption{Overall $\mathbb{G}$ scores of the tested optimization methods (left) and method success/failure in terms of $\mathbb{G}$ distribution (right). Bold bars on the left show average $\mathbb{G}$, while the pale rightward extensions show average $\mathbb{G}_{RW}$. On the right, the criteria for excellent and catastrophic performance were $\mathbb{G} > 0.9$ and $\mathbb{G} < 0$, with worst performance and best performance corresponding to achieving $\min \mathbb{G}$ and $\max \mathbb{G}$, respectively.}
	\label{fig:methods_G_performance}
\end{figure}

A simple, straightforward illustration of method performance may be attained by averaging $\mathbb{G}$ scores across the entire benchmark test, for every optimization method. This information can give some idea of method accuracy and is thus shown in Figure \ref{fig:methods_G_performance}. Considering that stochastic optimization application necessitates conducting multiple optimization runs, the "score bars" are extended with average $\mathbb{G}_{RW}$ values, which can be reasonably interpreted as statistically expected performance given 10 optimization runs.

These results make it rather clear which are the overall best and worst performing methods, as well as which methods can profit the most from repeated attempts at solving a problem. However, they do not show how the methods perform across different problem types, i.e. how often they exhibit particularly good or bad performance, or how robust they are.

So as to obtain some information on the distribution of $\mathbb{G}$ values, an additional analysis was conducted, where instances in which a method showed best, excellent, catastrophic, and worst performance, were separately counted. This data, also given in Figure \ref{fig:methods_G_performance}, allows for several observations. For example, methods such as PSO, ABC, and GA are rarely best of all, but often produce excellent results. On the other hand, methods like SSA, BA, and FWA are rather mediocre, rarely standing out one way or another. Interestingly, STOGO seems very much specialized, while LBFGSB is obviously entirely incapable for solving many of the highly multimodal problems, albeit still exceptionally accurate on a range of problems. It bears repeating that catastrophic performance stands for performance worse than pure random search, which makes a method like GWO look particularly bad, since it exhibits such performance on 22\% of test functions and yields worst results as often as it produces excellent results.

\subsection{Use-case specific method performance}

Estimating the characteristics or attributes of optimization problems in an objective and unbiased way is inherently difficult. However, when such estimates are available and since $\mathbb{G}$ values are normalized, method $\mathbb{G}$ for a given attribute can be computed as a weighted mean.
\begin{equation}
	\mathbb{G}^{[\text{att}]}_m = \frac{\sum_{p} \mathbb{G}_{p,m}\cdot w^{[\text{att}]}_p}{\sum_{p} w^{[\text{att}]}_p} ,
	\label{eq:att_G}
\end{equation}
where $[\text{att}]$ is the label of the observed attribute, $m$ and $p$ are method and problem indices, respectively, and $w_p$ is the weight of the same attribute for problem $p$.

In following analyses, we observe three complementary attributes pairs that can be objectively measured: global/local, fast/exhaustive and high-dimension/low-dimension. The complementarity of attribute pairs ($[\text{att+}]$, $[\text{att-}]$) rules the complementarity of their weights for each problem $p$:
\begin{equation*}
	w^{[\text{att+}]}_p + w^{[\text{att-}]}_p = 1.
\end{equation*}

Since LBFGSB is a pure local search method, we utilize its performance to determine weights for local and global (search) attributes. The local search weights are calculated as
\begin{align*}
	w^\text{[Loc.]}_p = \max\{0, \mathbb{G}_{\text{LBFGSB}, p}\},
\end{align*} 
for each problem  $p$, whilst complementary weight $w^\text{[Glob.]}_p=1-w^\text{[Loc.]}_p$.

The ability to deliver suitable results quickly (fast optimization), as well as after using the maximum number of evaluations (exhaustive optimization) are useful attributes of optimization methods. For these attributes, $\mathbb{G}_{p,m}$ in \eqref{eq:att_G} actually combines
relative metrics $\overline{\mathbb{G}}_{10\%}$, $\overline{\mathbb{G}}_{50\%}$, and $\overline{\mathbb{G}}_{100\%}$ for each problem, with respective weight factors $(1, 0.5, 0)$ for fast attribute and $(0, 0.5, 1)$ for exhaustive attribute.

Determining weights for problem dimensionality attribute is straight-forward. All problems are considered low-dimensional for $D\leq 10$, and high-dimensional if $D\geq 30$. In between, weights are linearly interpolated:
\begin{align*}
	w^\text{[Low $D$]}_p =
	\begin{cases}
		1 & \text{ if } D \leq 10,\\
		1 + (D - 10)/20 & \text{ if } 10 < D < 30,\\
		0 & \text{ if } D \geq 30.
	\end{cases}
\end{align*} 
Complementary high-dimensional attribute weights are determined as $w^\text{[High $D$]}_p=1-w^\text{[Low $D$]}_p$.

\begin{figure*}[htb!]
	\centering
	\includegraphics[width=1.0\linewidth, trim={0mm 0mm 0mm 15mm}, clip]{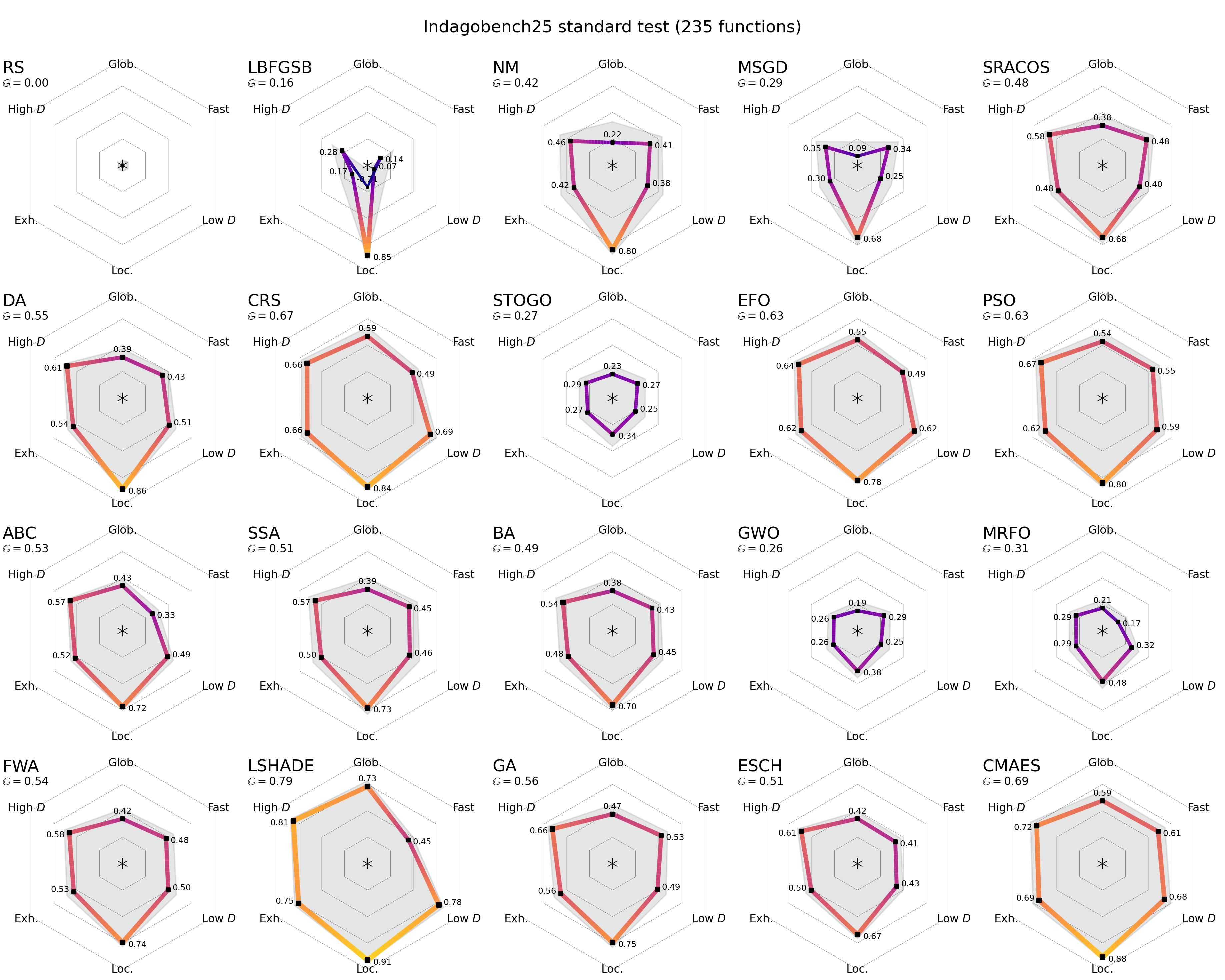}
	\caption{The visualization of assessed complementary attributes of analyzed optimization methods. While RS is the worst performing method (by design), it is interesting to observe that it surpasses LBFGSB in global search attribute. LSHADE outperforms all other methods across nearly all attributes except convergence speed, as its optimization process typically extends throughout the entire evaluation budget before stabilizing.}
	\label{fig:radar_plot_all_optimizers}
\end{figure*}

The above methodology allows us to estimate the attributes of optimization methods from the attributes of solved optimization problems. Three pairs of complementary attributes are suitably visualized for all methods in  Figure~\ref{fig:radar_plot_all_optimizers}.

\subsection{Sensitivity to problem multimodality}

Based on the benchmark test results, interesting information on method behavior can be extracted. One useful method property that can be assessed and analyzed is method versatility with regard to fitness function modality.

Noting that LBFGSB is a highly efficient, pure local gradient-based method, its performance can be taken as a kind of a posteriori measure of the effective multimodality of a test function at hand. In this practical premise, the better LBFGSB performs on a function, the closer to being effectively unimodal the function is. Based on this reasoning, a multimodality descriptor $M$ is introduced, calculated as:
\begin{align}
	M = \frac{1}{2} \left(1 - \mathbb{G}_{\text{\scriptsize LBFGSB}} \right) ,
	\label{eq:multimodality}
\end{align}
where $\mathbb{G}_{\text{\scriptsize LBFGSB}}$ is the result of LBFGSB on the used test function. For excellent performance of LBFGSB, i.e. $\mathbb{G}_{\text{\scriptsize LBFGSB}} = 1$, it is reasonable to assume that the function is unimodal, hence multimodality $M = 0$. On the other hand, in case of the method's extremely poor performance, i.e. $\mathbb{G}_{\text{\scriptsize LBFGSB}} = -1$, it makes sense to assume that the problem is highly multimodal, yielding $M = 1$. Consequently, $0 \ge M \ge 1$.

\begin{figure}[h]
	\centering
	\includegraphics[width=1\linewidth, trim={3mm 3mm 3mm 3mm}, clip]{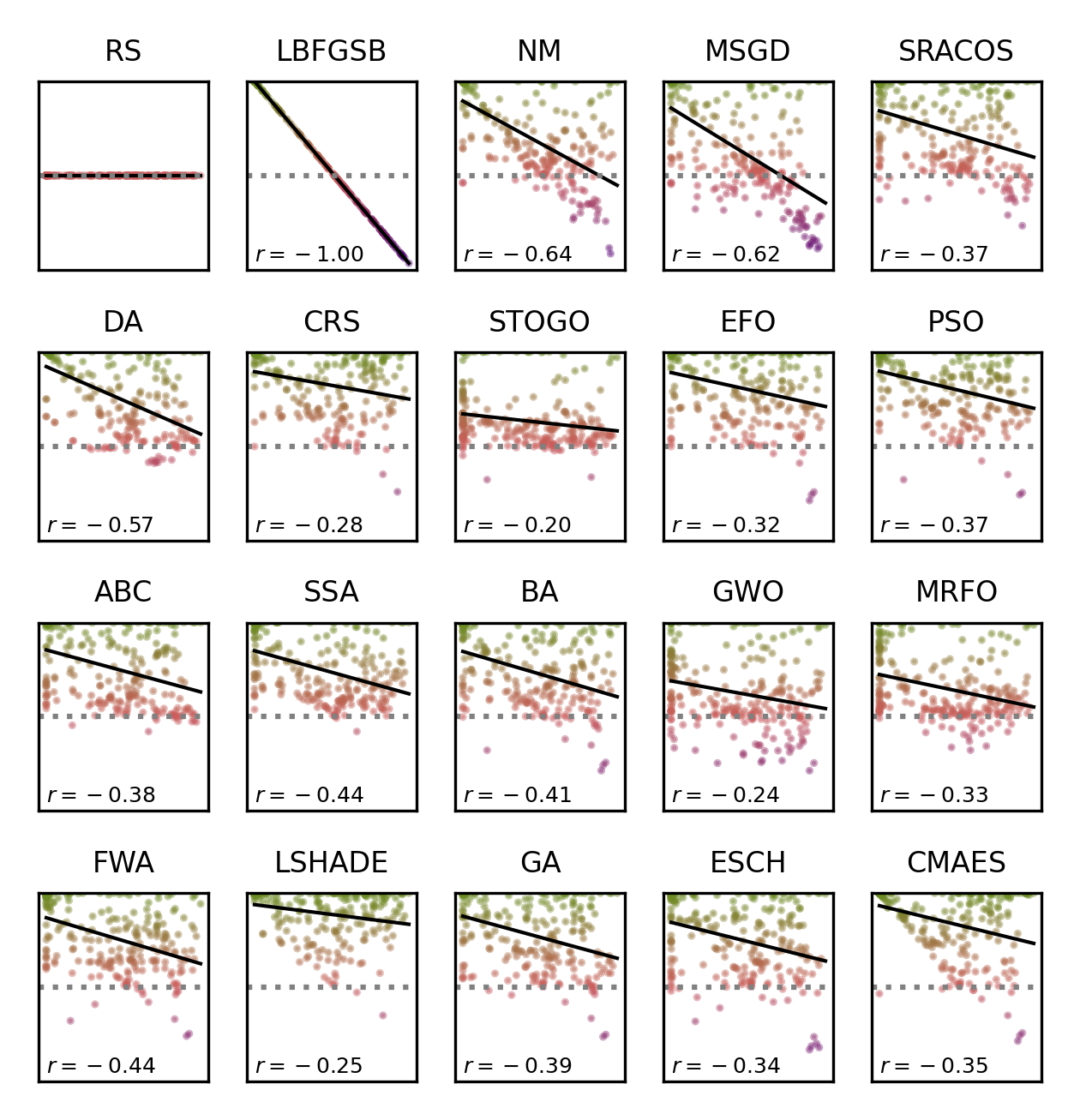}
	\caption{Optimization method versatility with regard to test function multimodality. Vertical axes represent $\mathbb{G}$ (higher is better) and horizontal axes represent $M$ (increasing rightward from 0 to 1); dotted lines mark $\mathbb{G} = 0$. Black lines are linear regressions; $r$-values are Pearson correlation coefficients. Note that, by definition of $M$ itself, LBFGSB necessarily yields perfect linear regression and $r=1$, and as such should not be analyzed in this context.}
	\label{fig:G_vs_M}
\end{figure}

For each analyzed optimization method, the obtained method performance scores $\mathbb{G}$ are related to the multimodality $M$ on each test function, and based on these $(M, \mathbb{G})$ data points simple regressions and correlations are computed so as to illustrate their relationship (Figure \ref{fig:G_vs_M}). These analyses can show the capabilities of each method for solving problems of increasing difficulty, i.e. they demonstrate the versatility of the optimization methods with regard to problem multimodality.

Based on these results, some modest observations can be made. As expected, it can be noted that all methods somewhat weaken with increased function multimodality, i.e. difficulty. Furthermore, the results clearly confirm that local methods NM and MSGD exhibit poorer performance on harder, strongly multimodal problems. On the other hand, it is obvious that some methods are more versatile than others, as CRS, STOGO, GWO, and LSHADE are the least prone to deteriorate when solving highly difficult problems.

\subsection{Computational complexity}

In order to assess the computational effort needed for an optimization run of a particular method, we can measure the time needed for the run to be carried out. This time can be defined as:
\begin{equation}
	T_t = T_m + n_{eval} \cdot T_f ,
\label{eq:total_time}
\end{equation}
where $T_t$ is the total time of an optimization run, $T_m$ is time taken by the optimization method algorithm, $T_f$ is time needed for one function evaluation, and $n_{eval}$ is the number of evaluations used. If we divide the equation \eqref{eq:total_time} by $n_{eval}$, we can formulate it as:
\begin{equation*}
	\frac{T_t}{n_{eval}} = T^{\circ}_m + T_f ,
\end{equation*}
where $T^{\circ}_m = T_m / n_{eval}$ is the computational time overhead of the optimization method per one function evaluation. By experimentally measuring $T_f$ and $T_t$ (and tracking $n_{eval}$), we can extract $T^{\circ}_m$. Note that $T^{\circ}_m$ represents the average of cumulative overhead across the complete optimization process, accounting for possible variations in overhead over iterations.

In order to remove the influence of the hardware and software platforms used, all measured $T^{\circ}_m$ times were normalized with regard to a reference time $T_{ref}$, for which the evaluation time of benchmark test function \texttt{CEC\_F1\_10D} was used. This enabled the introduction of a relative (dimensionless) computational complexity $C_m = T^{\circ}_m / T_{ref}$.

Method complexity $C_m$ was calculated for each optimization method and on each of the benchmark functions, and this data was used for comparing computational efficiency between different optimization methods. Execution time measurements were performed on a Microsoft Windows 10 machine with AMD Ryzen 9 7950X 16-core processor and the results of the analysis are given at the end of this section (Figure \ref{fig:method_features}).

\subsection{Method performance overlap}

For every particular optimization problem, there are optimization methods that are more suited and others which are less capable. However, it is reasonable to expect that some of the better optimization methods consistently outperform a number of other, worse performing methods.

In order to differentiate the specialized from the versatile methods, as well as detect the methods entirely superseded by others, an overlap in method performance was assessed. This was achieved by detecting the test functions on which a specific method was successful ($\mathbb{G} > 0.9$) and then determining the intersection of this "solved" function set with the corresponding "solved" function sets of all other methods. Method performance overlap was calculated as a share of functions of the method's "solved" functions set also appearing in another method's "solved" functions set.

\begin{figure}[h]
	\centering
	\includegraphics[width=1\linewidth, trim={3mm 3mm 3mm 3mm}, clip]{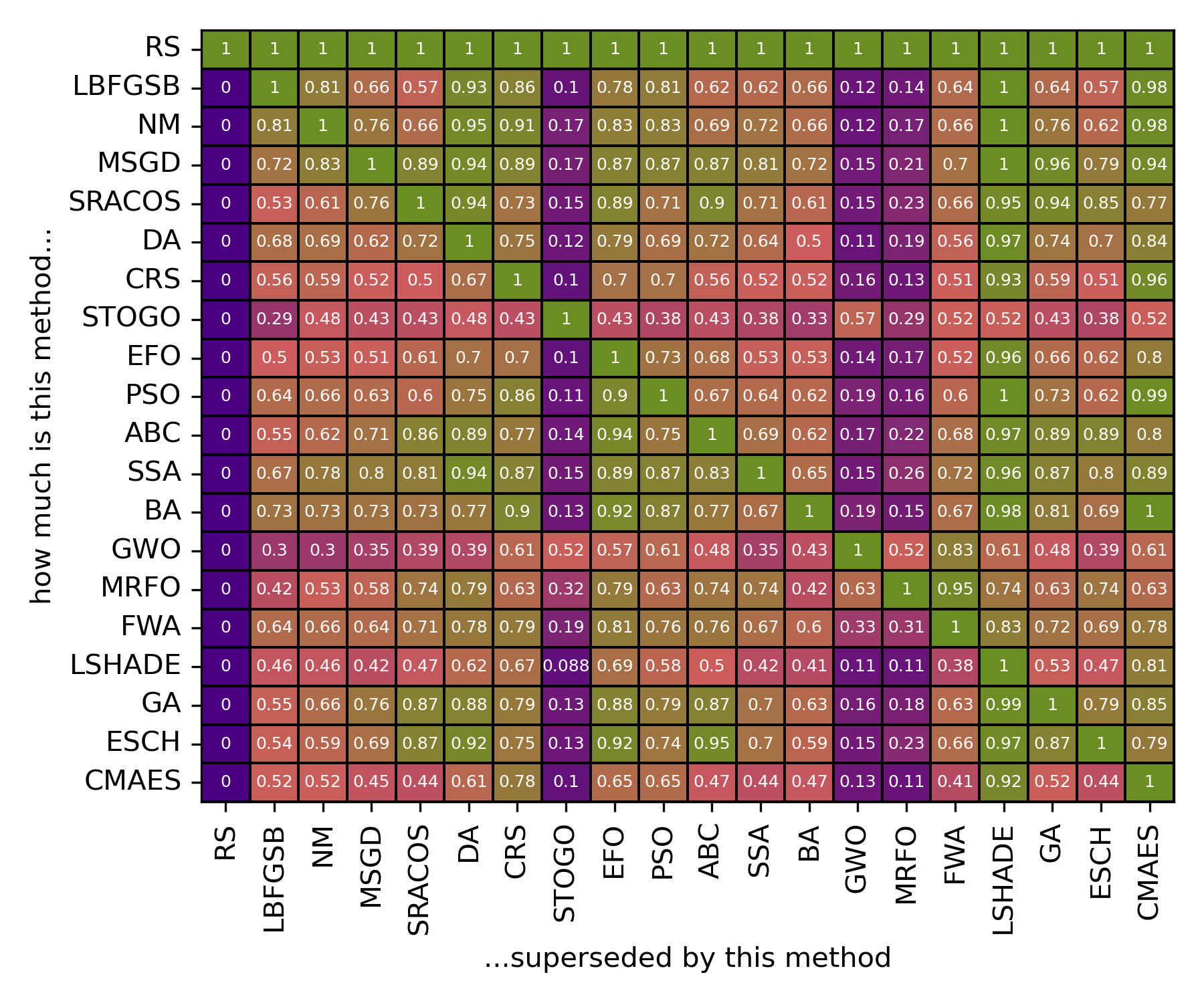}
	\caption{Shares of successfully solved problems ($\mathbb{G} > 0.9$) for a method (left) which are also successfully solved by another method (bottom). Less is better for a method in the corresponding row; more is better for it in its corresponding column. The numbers range from zero (none of the functions solved by the method on the left were also solved by a method on the bottom) to one (all of them are). For convenience, RS row values are set to one, although they are incalculable since $\mathbb{G}_{\text{\scriptsize RS}} = 0$, by definition.}
	\label{fig:overlap_heatmap}
\end{figure}

The results of this analysis shown in Figure \ref{fig:overlap_heatmap} illustrate the specialization and overall relative capabilities of the tested optimization methods. Some methods, such as STOGO and GWO, have their own particular niches in which they seem to be particularly well suited. On the other hand, some methods are strongly superseded by others, to the point that there is almost no reason for them to be used at all. For example, DA strongly supersedes (i.e. on $>90\%$ of functions) LBFGSB, NM, MSGD, SRACOS, SSA, and ESCH; CRS supersedes BA; EFO supersedes ABC, BA, and ESCH; ABC supersedes ESCH, FWA supersedes MRFO; GA supersedes MSGD and SRACOS. Similarly, CMAES performs well on 100\% of functions on which BA showed good results, and also strongly supersedes LBFGSB, NM, MSGD, CRS, and PSO. Finally, LSHADE supersedes all methods but STOGO, GWO, MRFO and FWA, while dominating LBFGSB, NM, MSGD, and PSO. It is also visible that PSO, otherwise a very competent method, does not strongly supersede any of the other tested methods.

\subsection{Best-performing method sets}

Information on the optimization methods' performance on the used benchmark test can be used to extract best-performing sets of methods. For this a combinatorial approach can be utilized, in which for all possible method combinations of fixed size the number of successfully solved fitness functions can be obtained. The best-performing method sets, with the methods given in the order of appearing in the sets of size $1, 2, \ldots 5$, are:
\begin{itemize}
	\item LSHADE, STOGO, CMAES, FWA, SRACOS (based on maximum number of problems solved with $\mathbb{G} > 0.9$)
	\item LSHADE, FWA, CMAES, STOGO, SRACOS (based on maximum number of problems solved with $\mathbb{G}_{RW} > 0.9$)
\end{itemize}
The problems solved with $\mathbb{G} > 0.9$ are here counted as "solved", and the problems with $\mathbb{G}_{RW} > 0.9$ are counted as "solved given 10 runs".

The results of this analysis provide some guidelines for choosing optimization methods. Namely, if only one optimization method is used, it should be LSHADE, which solved 54\% of the test functions over 10 optimization runs. If two methods can be afforded, than FWA and LSHADE should be used, which together can solve 58\% of the test functions. If five methods can be used, then it is (statistically) most opportune to choose LSHADE, FWA, CMAES, STOGO, and SRACOS, which together solved 65\% of the benchmark problems. STOGO, one of the weaker methods analyzed, is consistently featured in these best-performing sets. This leads to a conclusion that STOGO is very much a niche method, highly specialized in specific problems (as it can be also observed in Figure \ref{fig:overlap_heatmap}), which makes it very useful for inclusion in any kind of ensemble optimization strategy. This is true also for GWO, albeit to a lesser extent.

\subsection{Method properties}

\begin{figure*}[h!]
	\centering
	\includegraphics[width=1\linewidth, trim={3mm 3mm 3mm 3mm}, clip]{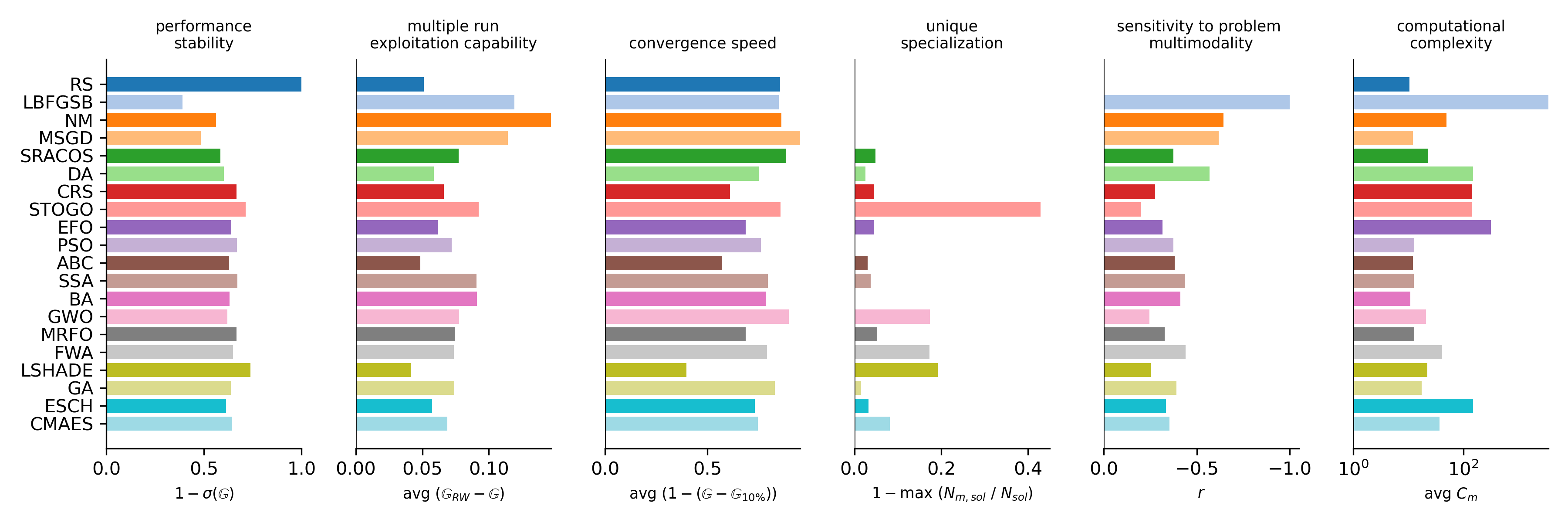}
	\caption{Summary of the optimization method properties, as assessed through the conducted analysis.}
	\label{fig:method_features}
\end{figure*}

Considering the significant scope and detail of the information obtained through the used benchmark test, several interesting method properties can be extracted. The compilation of various optimization method features is shown in Figure \ref{fig:method_features}.

Method properties given in Figure \ref{fig:method_features} are defined as follows:
\begin{itemize}
	\item \emph{Performance stability} serves as an indication of the variance of method accuracy over the entire benchmark test. It is calculated as $1 - \sigma \left(\mathbb{G}\right)$, i.e. as offset negative standard deviation of the performance grade.
	\item \emph{Multiple run exploitation capability} is a statistically expected increase of method accuracy if 10 optimization runs are used instead of one. It is defined as the difference between $\mathbb{G}_{RW}$ and $\mathbb{G}$, averaged across the entire benchmark test.
	\item \emph{Convergence speed} is based on the difference between $\mathbb{G}$ and $\mathbb{G}_{10\%}$ scores; the smaller the difference, the faster a method converges to a solution. It is calculated as $1 - \left( \mathbb{G} - \mathbb{G}_{10\%} \right)$ and then averaged across all functions.
	\item \emph{Unique specialization} is a measure of method irreplaceability in solving specific problems. It is defined as $1 - \max \left( N_{m, sol} / N_{sol} \right)$, where $N_{sol}$ represents the number of functions that the method in question successfully solved ($\mathbb{G} > 0.9$), while $N_{m, sol}$ represents the number of these same functions which are also successfully solved by another method $m$. This is all corresponding to the results given in Figure \ref{fig:overlap_heatmap}.
	\item \emph{Sensitivity to problem multimodality} is a quantitative measure of a rate of method performance deterioration with the increase of problem multimodality. Specifically, these are Pearson correlation coefficient $r$ numbers previously shown in Figure \ref{fig:G_vs_M}.
	\item \emph{Computational complexity} $C_m$ is measured and averaged across all non-simulation-based benchmark functions (namely CEC, AEP, ER, SP, EC, PP, and IPP). Simulation based problems were excluded due to their significant variation in function evaluation execution time (which is a consequence of using iterative solvers for the underlying mathematical models).
\end{itemize}

Finally, taking all this information, a number of most prominent optimization methods in terms of their exhibited behavior are emphasized in Table \ref{tab:MethodsHighlights}. It must be noted that these judgments should only be interpreted in the context of the conducted benchmark testing, i.e. for unconstrained real-variable engineering-type optimization problems and used software implementation.

\begin{table}[h]
	\centering
	\scriptsize
	\caption{Most prominent optimization methods}
	\begin{tabular}{>{\raggedright\arraybackslash}p{18mm} >{\raggedright\arraybackslash}p{29mm} >{\raggedright\arraybackslash}p{29mm}}
		Aspect & Top-ranked & Bottom-ranked \\
		\hline
		Overall accuracy & CRS, EFO, PSO, \textbf{LSHADE}, CMAES & \textbf{LBFGSB}, MSGD, STOGO, GWO, MRFO \\
		\hline
		Local search performance & \textbf{LBFGSB}, DA, CRS, LSHADE, CMAES & \textbf{STOGO}, GWO, MRFO \\
		\hline
		Global search performance & CRS, EFO, PSO, \textbf{LSHADE}, CMAES & \textbf{LBFGSB}, MSGD, STOGO, GWO, MRFO \\
		\hline
		High $D$ search performance & CRS, EFO, PSO, \textbf{LSHADE}, CMAES & \textbf{LBFGSB}, MSGD, STOGO, GWO, MRFO \\
		\hline
		Low $D$ search performance & CRS, EFO, PSO, \textbf{LSHADE}, CMAES & \textbf{LBFGSB}, MSGD, STOGO, GWO, MRFO \\
		\hline
		Fast search performance & SRACOS, PSO, FWA, GA, \textbf{CMAES}  & LBFGSB, STOGO, ABC, GWO, \textbf{MRFO} \\
		\hline
		Exhaustive search perform. & CRS, EFO, PSO, \textbf{LSHADE}, CMAES & \textbf{LBFGSB}, MSGD, STOGO, GWO, MRFO \\
		\hline
		Robustness & CRS, STOGO, PSO, SSA, \textbf{LSHADE} & \textbf{LBFGSB}, NM, MSGD, SRACOS, DA \\
		\hline
		Uniqueness & \textbf{STOGO}, GWO, FWA, LSHADE, CMAES & NM, MSGD, PSO, \textbf{BA}, GA \\
		\hline
		Repeated runs exploitation & LBFGSB, \textbf{NM}, MSGD, STOGO, SSA & CRS, EFO, ABC, \textbf{LSHADE}, ESCH  \\
		\hline
		Computational efficiency & MSGD, ABC, SSA, \textbf{BA}, MRFO & \textbf{LBFGSB}, DA, CRS, STOGO, EFO, ESCH \\
		\hline
	\multicolumn{3}{l}{\scriptsize \emph{Methods standing out the most are typed in boldface.}} \\
	\end{tabular}
	\label{tab:MethodsHighlights}
\end{table}

The often weak performance of local optimization methods (LBFGS, NM, MSGD) on functions which are mostly highly multimodal is entirely expected and should not be held against these methods. However, the application of local methods for solving global problems is not a rare occurrence in engineering practice, where the nature of the optimization problem at hand may be entirely unknown, and the given allowances of computational effort and work hours are strictly enforced. In such scenarios, local methods may often prove to be fully adequate, as is corroborated by the results of this benchmark test in which local methods perform competently on global optimization problems more often than some would expect. For example, consider NM and MSGD profiles in Figure \ref{fig:radar_plot_all_optimizers} or the fact that LBFGSB outperformed all other methods on 17\% of the test functions (Figure \ref{fig:methods_G_performance}), even often only marginally.

\section{Comparing $\mathbb{G}$ with alternative metrics}
\label{sec:cococompare}

In order to compare the behavior of $\mathbb{G}$-metric with existing alternative optimization performance rankings, several alternative metrics, typically used in similar analyses, were normalized and applied on the IndagoBench25 results.

One of these is the COCO analysis \cite{hansen2010real}, for which target fitness $f^\otimes$ values for all IndagoBench25 functions were defined as $f^\otimes = 0.1 f^+ + 0.9 f^-$ (which corresponds to $\rho = 0.9$, as per \eqref{eq:rho}). The second used metric is the Rank-based normalization \cite{rank_normalization}, and the third metric was the Min-Max normalization \cite{minmax_normalization}.

\begin{figure}[h]
	\centering
	\includegraphics[width=1\linewidth, trim={4mm 3mm 3mm 3mm}, clip]{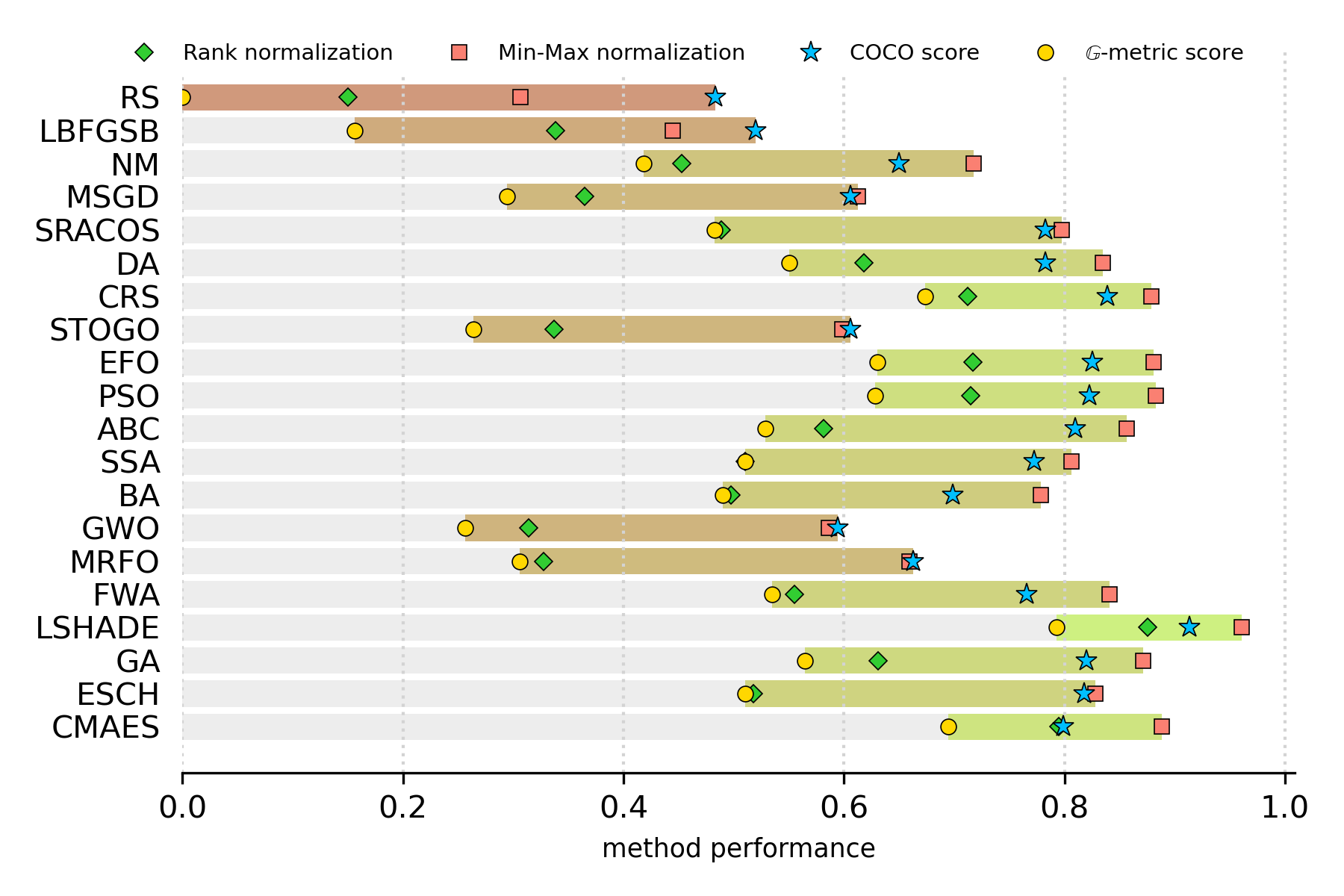}
	\caption{Comparison between several metrics on the IndagoBench25 testing results. Alternative metrics show good agreement with $\mathbb{G}$ scores, with the average std. dev. between the four metrics of $\sigma=0.13$. The best and worst-performing methods are similarly identified in all of the compared metrics, with the span across the metrics being the shortest for best-performing methods. Alternative metrics, most notably COCO, show a kind of overestimation of RS performance (relative to other methods), all the while tending to regard the majority of methods as approximately equally-well performing (Rank normalization $\sigma=0.18$, Min-Max normalization $\sigma=0.16$, COCO score $\sigma=0.12$, and $\mathbb{G}$ score $\sigma=0.19$). This allows for a problematic conclusion that most of the tested methods are performing just slightly better than RS.}
	\label{fig:COCOetc}
\end{figure}

The results of these metrics, together with the $\mathbb{G}$ scores, are given in Figure \ref{fig:COCOetc}, where it can be observed that $\mathbb{G}$-metric seems to yield better differentiation between methods, all without introducing subjective fitness thresholds such as $f^\otimes$, which gives it a significant advantage over other metrics.

\section{Supplementary material}
\label{sec:supplementary}

Comprehensive test results and accompanying materials for all 235 IndagoBench25 test functions are available online at \url{https://osf.io/7r6jz}. These include:
\begin{itemize}
	\item IndagoBench25 Benchmark Definition document
	\item Raw benchmark test results (\emph{IndagoBench25-R1} dataset)
	\item Convergence plots for all IndagoBench25 functions
	\item IndagoBench25 user instructions
	\item IndagoBench25 problems source code
	\item IndagoBench25 benchmark methodology source code
	\item Partial benchmark test results for TPE
\end{itemize}
All these can be freely used for any future analyses or independent benchmarking.

\section{Conclusion}


In this paper we propose a methodology for optimization evaluation which relies on random sampling as a references for the performance metric. The developed metric is utilized with IndagoBench25, a benchmark suite of 235 continuous bounded unconstrained optimization problems, most of which are engineering problems. The proposed methodology was applied for benchmarking 20 optimization methods in a comprehensive, statistically sound manner. Each method was tested on each test function at least a 100 and up to a 1000 times, following the statistical convergence requirements.

The utilized $\mathbb{G}$-metric has shown to be robust and very informative, while providing a fair evaluation on real-world method performance. Thus it has revealed that even some of the well-regarded optimization methods can experience a total performance collapse (i.e.  exhibit worse performance than random search) on particular problems. Furthermore, the conducted analysis has confirmed what is well known: that every method has its area of competence (no matter how narrow a niche it may be), and no method is universally better than all other methods across all possible problems \cite{wolpert2002no}. However, the analysis has also clearly demonstrated that some optimization methods fail to confidently prove their worth, while some other methods positively supersede them. 

Besides IndagoBench25 and the $\mathbb{G}$-metric, a post hoc technique for assessing the multimodality of an optimization function is proposed. This technique is effectively employing a local gradient-based method (LBFGSB) as an exploratory tool for measuring optimization problem globality.

The conducted analyses allow for several specific remarks regarding tested optimization methods' performance:
\begin{itemize}
	\item LSHADE is the overall best performing method, followed by CMAES, and to a degree CRS, EFO and PSO.
	\item The best methods are performing very well across almost all metrics and use cases. Therefore it is advised to use three or four excellent methods rather than employ many methods of various specialization in the hope of stumbling onto "the right one".
	\item Some methods (e.g. STOGO, GWO) appear to have their own niche specializations and as such could possibly contribute in a wider analysis, if affordable.
	\item While they are generally less competitive against stochastic algorithms in global search tasks, local methods (LBFGSB, NM, MSGD) can sometimes outperform certain global methods in specific optimization scenarios. 
\end{itemize}

The proposed methodology certainly has some possible shortcomings. 
Obtaining the reference point $f^-$ depends on the thoroughness of the conducted optimizations and utilized methods, and there is no guarantee of how precisely it is determined, especially for black-box problems.
It could be argued that using median for tracking convergence is problematic, thus other statistical approaches should be investigated. Furthermore, the $\mathbb{G}$-metric becomes unstable when an optimization results with $\mathbb{G} < -1$ (i.e. $f > f^+$), albeit this should manifest only on severely degraded optimization procedures. 

There remains the question of the used test suite adequacy. IndagoBench25 could hide some prominent biases, or could simply be insufficiently representational for a generic engineering optimization test. Future research should investigate this further, potentially leading to an updated version of IndagoBench designed to address these limitations.

Nevertheless, the proposed benchmark metric and test function suite could be used to test as many of the most popular optimization methods as possible, which would help in assessing the practical aspects of the ever-growing landscape of optimization algorithm research. Furthermore, the presented benchmark procedure could be complemented with Exploratory Landscape Analysis techniques to provide additional perspectives. Moreover, even the proposed benchmarking procedure itself, administered on a selection of multifarious optimization methods, could be used as a tool for investigating the properties of optimization problems.

\section*{Acknowledgments}
This research was supported by University of Rijeka grants \emph{uniri-iskusni-tehnic-23-52} and \emph{uniri-mzi-25-38}, and University of Rijeka Center for Advanced Computing and Modeling.

\bibliographystyle{IEEEtran}  
\bibliography{bibliography}

\vfill

\end{document}